\documentclass [10pt,twocolumn]{IEEEtran}
\usepackage{algorithm,algorithmic,amsbsy,amsmath,amssymb,epsfig,bbm,mathrsfs,multirow,amsthm}
\usepackage[T1]{fontenc}
\usepackage[latin9]{inputenc}
\usepackage{tabularx}
\usepackage{longtable}
\newcolumntype{L}[1]{>{\raggedright\arraybackslash}p{#1}} 
\newcolumntype{C}[1]{>{\centering\arraybackslash}p{#1}}
\newcolumntype{R}[1]{>{\raggedleft\arraybackslash}p{#1}}

\usepackage{array, tabularx, boldline}
\usepackage{eurosym}
\usepackage{amstext} 
\DeclareRobustCommand{\officialeuro}{%
	\ifmmode\expandafter\text\fi
	{\fontencoding{U}\fontfamily{eurosym}\selectfont e}}

\usepackage{subcaption}

\begin{document}
	
	\title{\huge Computation Offloading and Content Caching Delivery in Vehicular Edge Computing: A Survey}
	
	\author{Rudzidatul Akmam Dziyauddin, \textit{Senior Member, IEEE}, Dusit Niyato, \textit{Fellow, IEEE}, Nguyen Cong Luong, Mohd Azri Mohd Izhar, \textit{Senior Member, IEEE}, Marwan Hadhari, \textit{Senior Member, IEEE}, Salwani Daud, \textit{Senior Member, IEEE}
	\thanks{R.~A.~ Dziyauddin, Mohd Azri Mohd Izhar dan Salwani Daud are with Razak Faculty of Technology and Informatics, Universiti Teknologi Malaysia, Malaysia. E-mail: rudzidatul.kl@utm.my}
	\thanks{D.~Niyato and N.~C.~Luong are with School of Computer Science and Engineering, Nanyang Technological University, Singapore. E-mails: dniyato@ntu.edu.sg., clnguyen@ntu.edu.sg}
	\thanks{Marwan Hadri Azmi is with School of Electrical Engineering, Universiti Teknologi Malaysia. e-mail:hadri@utm.my}}
	
	\maketitle
	\begin{abstract}
	
	Autonomous Vehicles (AVs) generated a plethora of data prior to support various vehicle applications. Thus, a big storage and high computation platform is necessary, and this is possible with the presence of Cloud Computing (CC). However, the computation for vehicular networks at the cloud computing suffers from several drawbacks, such as latency and cost, due to the proximity issue. As a solution, the computing capability has been recently proposed at the edge of vehicular networks, which is known as Vehicle Edge Computing (VEC). This leads to other open problems for vehicles to offload and compute data at edge nodes, and also how data is cached in edge nodes and then disseminated to other vehicles. In this paper, we initially present an	overview of VEC architectures including types of layers, fog nodes, communication technologies and also vehicle applications, which are used in data offloading and dissemination scenarios. Since the mobility is critical on the VEC performance, the mobility model used in the VEC scenario is also discussed. We extensively review the Computation Offloading (ComOf) techniques as well as Content Caching and Delivery (CachDel) approaches for VEC. We finally highlight some key research challenges, issues and future works in the paper.
		
	\end{abstract}

\section {Introduction} 

Recent developments in the edge computing field have led to a renewed interest in vehicular edge networks. With the increasing number of AVs on the road, the computation at the cloud platform becomes a great of concern. Each AV is envisaged to generate 1 GB data for every second \cite{shi2016edge} and on average 30 TB data in a day. Any AV relying on the cloud \cite{gerla2012vehicular} for data computation might crash because of millions of vehicles transmitting and receiving immense of data from a data center for processing. It is even crucial with the emerging of smart vehicles equipped with a massive number of sensors and human computer interaction devices for supporting intelligent traffic and navigation applications, such as active driving safety assistance, smart parking and live traffic management \cite{Zhang2017,7981532}. Due to the proximity issue of CC to vehicles, CC suffers from serious shortcomings such as latency, high overhead and efficiency, particularly for time-sensitive vehicle applications \cite{8370877}, \cite{Dai2018}. Therefore, fast computation and decision making are necessary to deliver short response times, dynamic processing and low dependency on the cloud networks. This is viable when the intelligence and processing capabilities are pushed down closer to where the data originates that leveraged on nearby vehicles, Road Side Units (RSUs), Base Stations (BSs) and Mobile Edge Computing (MEC) servers known as edge nodes. This type of computing is called Vehicular Edge Computing. 



The VEC is feasible since AVs have computing power of 106 Dhrystone Million Instructions Executed per Second (DMIPS) in the near future \cite{Intel}, which is tens times of the current laptops that make cars called as computers-on-wheels. Although AVs are generally equipped with onboard units (OBUs), the computing and storage capability is only on small-scale that make them  dependent on other computational resources. Some on-board multimedia applications require stringent deadline and high computation, particularly the real-time applications, thus the limited computing and storage resources of vehicles are hard to support computation-intensive applications. The OBUs can perform simple computations, collect local data from sensing devices, and upload data to the edge nodes either stationary or mobile. By extending CC to the edge of the networks, the stationary edge node, such as RSUs, BSs and MEC servers can provide high reliability, high bandwidth, and low-latency computing services for the requesting vehicles. This greatly reduces the communication delay and avoids congestion. In addition, the MEC server can obtain the user's surrounding environment information in real-time and hence it can optimise the services dynamically and rapidly. Another feasible solution is for AVs to offload its data to other nearby AVs with cheap payments, but with a limited storage. Enabling cooperative driving among AVs, such as platoon-based driving \cite{HPeng2017,HPengDLi2017}, convoy-based driving \cite{ZSuYhui2018}, Internet of Vehicles (IoV) \cite{zhang2019mobile} and Vehicular Social Networks (VSN) \cite{8517127} can aid VEC becomes possible.


The key issue of VEC is the computation offloading mechanism whereby the vehicles must select optimum edge nodes in-real time by satisfying the latency requirement, low cost and high energy efficiency. Also, with such schemes the service provider can still gain some profits. Another critical issue is  where the contents are cached at what edge nodes and then delivered directly to the corresponding vehicles. Much uncertainty still exists about the data computation offloading techniques and also content caching and delivery, i.e., downloading that are closely related to the optimisation problem of resource management in VEC. From the  literature, ComOf and CachDel mostly involve with the optimisation of scheduling and task allocation mechanisms, which are the primary scope of our survey paper. 

Existing surveys generally discussed the mobile data offloading technologies \cite{zhou2018survey} particularly in cellular networks \cite{chen2015survey}, mobile edge computing \cite{mach2017survey}, opportunistic offloading \cite{xu2018survey} and game theory in multi-access edge computing \cite{moura2019survey}. However, very little attention has been paid to the VEC \cite{raza2019survey}. Since the ComOf and CachDel mechanisms are of paramount importance to both VEC and VFC, this motivates us to fill in the gap by conducting a comprehensive survey on both offloading and downloading scenarios. The contributions of our survey work in this field are threefold. First, from a
comprehensive literature surveys, the general VEC architecture is defined in terms of its layer,
the types of communication technology, and vehicle applications. Second, our work summarises existing VEC architectures in terms of layers, fog nodes, communication technology, vehicle applications as well as the mobility model used. Third, we reviews the optimization techniques for ComOf and CachDel problems for VEC. 

\begin{figure*}[]
	\center{\includegraphics[width=1\textwidth]	{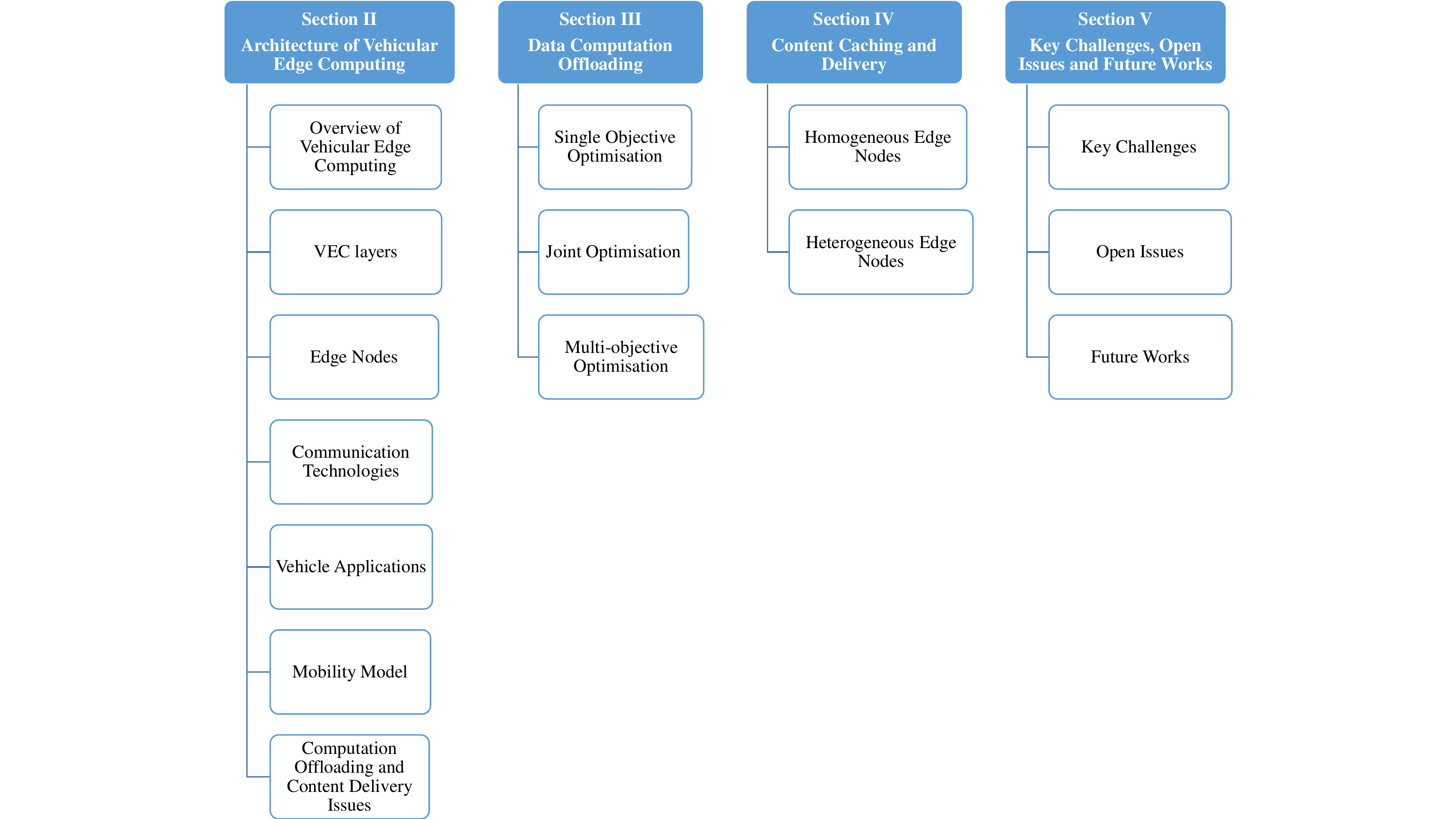}}
	\caption{A taxanomy of the survey on ComOf and CachDel for VEC}
	\label{fig:outline}
\end{figure*}

Figure \ref{fig:outline} illustrates the taxanomy of this paper. The rest of the paper is organised as follows. Section \ref{sec:architecture} explains a general VEC architecture as well as existing VEC architectures that consist of VEC layers, edge nodes, communication technologies, vehicle applications and mobility model used. The computation offloading and content delivery issues are also discussed in the section. Section \ref{sec:ComOf} discusses the data computation offloading solutions in three different groups, which are single objective optimisation, hybrid optimisation, and multi-objective optimisation. Section \ref{sec:CachDel} presents content caching and delivery approaches for VEC in two groups, which are homogeneous and heterogeneous edge nodes. Section \ref{sec:issues} highlights the key challenges, open issues and future works. The paper is concluded in Section \ref{sec:conclusion}. The abbreviations used in the paper are listed in Table \ref{fig:abb}.

\begin{table}[]
	\caption{List of abbreviations}
	\label{fig:abb}
	\begin{tabular}{|c|l|}
		\hline
		Abbreviation & \multicolumn{1}{c|}{Definition}                    \\ \hline
		AV           & Autonomous Vehicles                                \\ \hline
		BS           & Base Stations                                      \\ \hline
		CachDel      & Content Caching and Delivery                       \\ \hline
		CC           & Cloud Computing                                    \\ \hline
		ComOf        & Computation Offloading                             \\ \hline
		DMIPS        & Dhrystone Million Instructions executed Per Second \\ \hline
		GPS          & Global Positioning System                          \\ \hline
		ICN          & Information Centric Networking                     \\ \hline
		I2I          & Infrastructure-to-Infrastructure                   \\ \hline
		I2V          & Infrastructure-to-Vehicle                          \\ \hline
		IoV          & Internet of Vehicles                               \\ \hline
		ITS          & Intelligent Transport System                       \\ \hline
		MEC          & Mobile Edge Computing                              \\ \hline
		NFV          & Network Function Virtualisation                    \\ \hline
		OBUs         & Onboard Units                                      \\ \hline
		PV           & Parked Vehicles                                    \\ \hline
		QoS          & Quality of Service                                 \\ \hline
		RSU          & Road Side Units                                    \\ \hline
		SDN          & Software-Defined Networks                          \\ \hline
		SEN          & Stationary Edge Nodes                              \\ \hline
		V2I          & Vehicle-to-Infrastructure                          \\ \hline
		V2V          & Vehicle-to-Vehicle                                 \\ \hline
		VEC          & Vehicular Edge Computing                           \\ \hline
		VFN          & Vehicular Fog Node                                 \\ \hline
		VM           & Virtual Machine                                    \\ \hline
		VSN          & Vehicular Social Networks                          \\ \hline
	\end{tabular}
\end{table}

\section{Architecture of vehicular edge computing}
\label{sec:architecture}
To realise edge computing in a vehicular environment, the architecture of VEC is essential as it dictates the granularity of the resource management algorithms. This section gives an overview of VEC and then discusses the VEC architecture in terms of types of layers, edge nodes and communication technologies employed. The mobility models considered for the vehicles are also highlighted in this section.

\subsection{Overview of vehicular edge computing}
The three-layer of a VEC architecture, namely, the cloud layer, edge layer and smart vehicular layer in \cite{raza2019survey} is similar to the general edge computing architecture in \cite{hu2017survey}. Even if the survey includes the three VEC layers \cite{raza2019survey, hu2017survey}, our survey, in order to focus on the ComOf and CachDel, defines more detail VEC layers where the edge cloud layer is further divided into vehicle edge nodes layer and static edge nodes layer. We then expand the three-layer into four-layer whereby the nodes in the edge layer are categorised into vehicle edge nodes and stationary edge nodes layers.

\begin{figure*}[h]
	\center{\includegraphics[width=0.7\textwidth]	{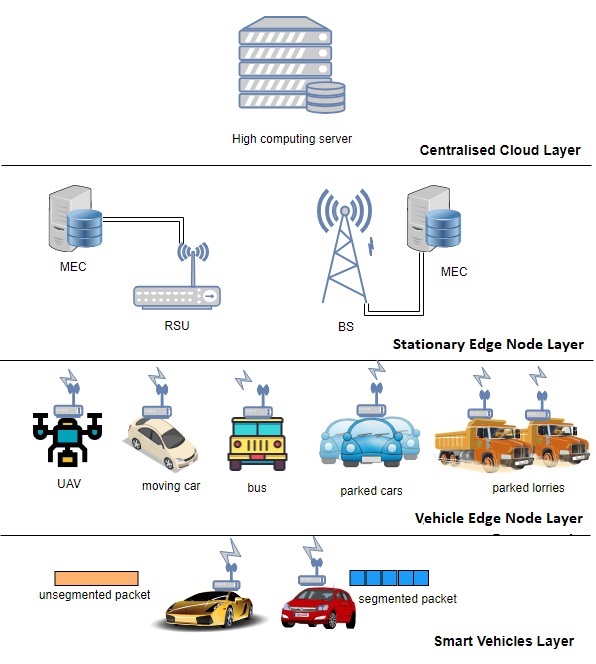}}
	\caption{A general four-layer VEC architecture}
	\label{fig:VEC-layer}
\end{figure*}

Figure \ref{fig:VEC-layer} illustrates a general four-tier VFC architecture resumed from the bottom layer, namely, smart vehicles, vehicle edge nodes, stationary edge nodes, and the top layer that is a centralised cloud. The layers are briefly explained as follows.  \\

\begin{itemize} 
	\item \textbf{Smart Vehicles} 
		Any types of vehicles that can be autonomous and equipped with OBUs\cite{raza2019survey, lopez2017internet, olariu2011taking, abuelela2010taking}, number of sensors which includes Radio Detection and Ranging (RADAR) and Light Detection and Ranging (LIDAR), Global Positioning System (GPS), videos and cameras. The OBU has features of computation, storage and also networking \cite{raza2019survey}. The smart vehicles are sometimes known as service requestors \cite{LiWang2018}, client vehicles \cite{CZhu_Conf2018, CZhu_2018}, task vehicles \cite{YSunPro2018, YSun2019} and AVs \cite{8370877, AVE2017, HMenouar2017, peng2019spectrum} in the VEC architecture. The vehicles under this layer can upload or download segmented/unsegmented data to or from nearby edge nodes. The former is often described as offloading, and the latter is known as the content delivery or dissemination. \\
	
	\item \textbf{Vehicular edge nodes} 
	Smart vehicles with available resources can establish a vehicular cloudlet and share their resources to the requesting smart vehicles within their coverage. The vehicle edge node can be considered to be a vehicle service provider \cite{JKang2019} or a service vehicle \cite{YSunPro2018, YSun2019}. Note that the vehicles can either be in moving \cite{CZhu_2018} or parking states \cite{YZhang2018, YZhang2019}.\\
	
	\item \textbf{Stationary edge nodes}
	An RSU or a typical cellular BS is connected to an MEC server for a high computation and storage capability. The MEC server can be co-located at the BS \cite{QYuan2018, qi2018vehicular} or RSU \cite{YSunPro2018, JDu2019}. However, the cost can be reduced if a pool of RSUs can share the MEC server\cite{Zhang2017, li2019compound}. The RSU is more preferable than the BS to serve the vehicles due to its proximity to smart vehicles and mobile edge nodes \cite{luo2018cooperative}. In case of that there is no RSU coverage, the BS can still provide the edge computational service.\\
	
	\item \textbf{Centralised cloud}
	A centralised cloud \cite{8466353, sym10110594} consists of a large number of dedicated servers with great computation, storage capability, and also stable connections. Nevertheless, the trade-off is a high cost of computational service and can cause a long delay.\\
\end{itemize} 

In the next section, we will describe VEC layers that are used in the existing ComOf and CachDel schemes.

\subsection{VEC layers}
In general, the VEC architecture may consist of different number of tiers, which are two, three or four. Table \ref{type-layers} tabulates some examples of VEC architectures in terms of layers, name of the VEC systems, their advantages and disadvantages. 

\begin{table*}[!h]
	\centering
	\caption{A summary of VEC architectures}
	\label{type-layers}
	\begin{tabular}{|C{1.5cm}|C{4cm}|L{3.5cm}|L{3.5cm}|L{3.5cm}|}\hline
		\textbf{Related Works}      & \textbf{Name of Layers/planes} & \textbf{Name of System} &\textbf{Advantages} &\textbf{Disadvantages}\\  \hline
		
		\cite{8535091}	 &\begin{minipage}[t] {0.2\textwidth} \begin{itemize} \item Computing domain  (vehicular cloud computing and vehicular edge computing) \item Data domain \end{itemize} \end{minipage}     & Big data enabled EnerGy-efficient vehIcular edge computing (BEGIN)  &big data analytics &high computing platform and high cost  \\ \hline
		
		\cite{8539726}	    &\begin{minipage}[t] {0.2\textwidth} \begin{itemize} \item Data Plane	\item Control Plane   \end{itemize} \end{minipage}  & Software-Defined Vehicular Edge Computing (SD-VEC) & SDN controls resource management for high efficiency &high overhead for control messages \\   \hline
		
		\cite{8517127}	       &\begin{minipage}[t] {0.2\textwidth} \begin{itemize} \item Vehicular social edge computing  \item End user \end{itemize} \end{minipage}  & Vehicular Vocial Edge Computing (VSEC) &vehicular social edge computing layer satisfies QoE &content catching can reduce storage of vehicles\\ \hline
		
		\cite{wang2018contract, Zhou2019, QYuan2018, zhou2018joint, CYang2019, QLiu2018, JDu2019, FSun2018, li2017resource, SMu2018,YSunPro2018,YSun2019}     &\begin{minipage}[t] {0.2\textwidth} \begin{itemize} \item Edge cloud \item Vehicles \end{itemize} \end{minipage}      & Vehicular Edge Computing Networks (VECN) \cite{CYang2019} , Fog-enhanced Radio Access Networks (FeRANs) \cite{li2017resource}                     & low delay and overhead       &  insufficient resources for high density areas  \\    \hline
		
		\cite{8466353, 8370877,midya2018multi, ZNing2019, el2019exploiting}  &\begin{minipage}[t] {0.2\textwidth} \begin{itemize} \item Vehicles, access \cite{8466353} \item Edge \item Central or core cloud \end{itemize} \end{minipage}   & Software-Defined and FC-based Vehicular NETwork (SDFC-VeNET) \cite{8466353}      &additional resources or resource management controller   & high overhead and long delay               \\    \hline
		
		\cite{JWang2018}	 &\begin{minipage}[t] {0.2\textwidth} \begin{itemize}   \item Smart Pervasive Service layer (L-SPS)    \item Dynamic resource adaption layer (L-DRA)  \item Collaborative network component layer (L-CNC) \end{itemize} \end{minipage}    &Smart identifier network(SINET) & constitutes of function-node autonomic mapping and service-function autonomic mapping  & high complexity for real implementation    \\ \hline
		
		\cite{GQiao2018}  &\begin{minipage}[t] {0.2\textwidth} \begin{itemize}   \item Cloud-enabled control layer \item Mobile edge computing layer \item Multi-access connected cloudlet layer \end{itemize} \end{minipage}	&Vehicular Edge Multi-Access Network (VE-MAN) &cloud-enabled control layer has a global view on traffic environment and network state  & vehicular cloudlet leader might change at certain period that can interrupt the offloading and computing service  \\ \hline
		
		\cite{8434345}    &\begin{minipage}[t] {0.2\textwidth}\begin{itemize} \item Application \item Control(i.e. SDN Controller) \item Transmission, caching and computation \item Road and vehicular plane   \end{itemize} \end{minipage}		    &Software-defined vehicle networks with MEC  & SDN controller    & no vehicles as edge nodes\\	\hline

		\cite{8094861}     &\begin{minipage}[t] {0.2\textwidth} \begin{itemize}   \item Cloud  \item Local Authority and VEC Servers  \item Network Infrastructure (RSU and BS) \item Road and vehicular plane \end{itemize} \end{minipage}   &Distributed REputAtion Management System (DREAMS) & local authority handles reputation of vehicles, monitors networks, record information and update blacklist     & high overhead with the reputation information           \\  \hline
		
		\cite{8522034}      &\begin{minipage}[t] {0.2\textwidth} \begin{itemize} \item Requesting vehicles \item Service provider  \item VEC Servers  \item Parked vehicles   \end{itemize} \end{minipage}	 	& Parked Vehicle Edge Computing (PVEC) & service provider divides task into multiple subtasks, selects parked vehicles for computing and manages the reward    & high overhead                     \\  \hline

	\end{tabular}
\end{table*}

Mostly ComOf works focused on the two primary layers, namely the vehicle and edge layers \cite{Zhou2019, QYuan2018, zhou2018joint}, which can be distinguished in terms of types of edge nodes used, as discussed in the next subsection.  However, some works introduced different types of two tiers, such as Big data enabled EnerGy-efficient vehIcular edge computing (BEGIN) \cite{8535091} proposed a computing domain and a data domain whilst Software-Defined Vehicular Edge Computing (SD-VEC) \cite{8539726} proposed a data plane and a control plane in their VEC architectures. The two layers can also be in the form of types of computing, cloud and edge computing layers \cite{WZhang2017,peng2018sdn}.

The third layer of the VEC architecture involves with the cloud, such as core or central cloud \cite{8466353,8370877,midya2018multi}, cloud computing \cite{FLin2018, XChen2017},  cloud-enabled control layer \cite{GQiao2018}, regional cloud \cite{sym10110594}, public cloud \cite{LPu2019, JKang2019}, cloud server layer \cite{el2019exploiting}, remote cloud \cite{Pham_2019}. The key advantage of the third layer is to provide an additional computational resources in the case of insufficient resources at edge nodes. Unlike others, the proposed VEC layers in \cite{JWang2018} constitutes of three unique layers as seen in Table \ref{type-layers}. 

Subsequently, the application layer \cite{8434345}, local authority \cite{8094861}, city-wide controller \cite{luo2018cooperative} or service provider\cite{8522034} is the fourth and also the last layer expanded in the VEC architecture. The entities in such layer deal with the Quality of Service (QoS)\cite{8434345}, resource allocation \cite{luo2018cooperative} or reward policy \cite{8094861, 8522034}. However, the bottleneck is the high overhead as the number of control messages, for instance, the amount of storage offered and computation time, may presence in the system. Regardless number of tiers in the VEC, interestingly for the scenario of vehicles social communication a layer related to the social edge computing \cite{8517127, zhou2018social, LZhangZhao2018} is introduced. The social relationships of vehicles are created based on their social interest and ties \cite{LZhangZhao2018}, preferences of content selection \cite{zhou2018social}, or temporary storage of current contents or movies\cite{8517127}.

Apart from that, the enabler technology, such as big data \cite{8535091}, blockchain \cite{JKang2019},  Software-Defined Networks (SDN) \cite{8539726,huang2018v2v, JLiu2017,DAChekired2018} and Network Function Virtualisation (NFV) \cite{JWang2018, peng2019spectrum} are also incorporated in the VEC system.  The SDN and NFV bring new insights as their benefits can be in the functionalities presented in Table \ref{sdn-table} and Table \ref{nfv-table}, respectively. Typically, SDN is used to separate the control plane and data plane in serving VEC \cite{8539726, peng2018sdn} and even to separate the control plane for cloud and fog servers \cite{DAChekired2018}. Since SDN stored the contextual information of vehicles \cite{huang2018v2v, JLiu2017}, such as vehicle identification, location, speed, link, and contents in a database that resulted to an optimal offloading decision including vehicle-to-vehicle (V2V) paths and handover \cite{huang2018v2v}. With the information and optimal decision, the performance of a VEC system can then be predicted \cite{JLiu2017}. In addition to that, the resource abstraction and slicing \cite{8535091} as well as wireless access interworking \cite{peng2018sdn} are handled by SDN. Therefore, SDN has a potential to improve the agility, reliability, scalability, and latency performance in VEC \cite{liu2017scalable}.
 
On the other hand, NFV primarily concentrates on the creation and configuration of virtual resources whereby the task is variably segmented dependent on the available virtual machine (VM) \cite{WZhang2017, JWang2018}. Despite that, the spectrum slicing, load balancing and power control can also be administered by NFV \cite{peng2018sdn,peng2019spectrum}. Therefore, NFV offers scalable virtual resources for scheduling different tasks in a timely manner. The connected edge server including the SDN/NFV controller offers important capabilities for identifying and selecting edge nodes, allocating and migrating tasks \cite{li2017resource} and giving rewards to the AV that undertakes the residual workloads \cite{sym10110594}.

%

\begin{table*}[!h]
	\centering
	\caption{Functionalities of SDN in VEC}
	\label{sdn-table}
	\begin{tabular}{|c|l|} \hline
		\textbf{Related Works}  &\textbf{Functions of SDN}                                                                                                                                                    \\ \hline
		\cite{8535091}    &\begin{minipage}[t] {0.5\textwidth} \begin{itemize}                                                                                                                           \item abstract and slice physical infrastructure resources into distinct virtual vehicular networks \item administers the complex control and management functionalities \end{itemize} \end{minipage}    \\ \hline                               
		
		\cite{8539726} &\begin{minipage}[t] {0.5\textwidth} \begin{itemize}                                                                                                                           \item  facilitates flexible and dynamic network management by separating control plane and data plane  \item collects all the data plane information periodically \item facilitates optimal task offloading  \end{itemize} \end{minipage}    \\ \hline                               
		
		\cite{peng2018sdn} &\begin{minipage}[t] {0.5\textwidth} \begin{itemize}    \item splits and handles control plane and data plane. \item wireless access interworking \item abstracts and reallocates diverse radio spectrum resources to BSs  \end{itemize} \end{minipage}    \\ \hline
		
		\cite{8466353}    &\begin{minipage}[t] {0.5\textwidth} \begin{itemize}    \item provide control functions, including radio resource management, mobility management, communication management, traffic management \item receives real-time vehicle information, such as speed, traffic density, channel state information (CSI) and queue state information (QSI) \end{itemize} \end{minipage}    \\ \hline     
		
		\cite{DAChekired2018} &\begin{minipage}[t] {0.5\textwidth} \begin{itemize}    \item  collects information from vehicles, RSUs, and BSs and servers within the fog cell (information-gathering module). \item manages different wireless networks (wireless vehicles network manager module). \item generates control directives and forwards them to all devices in the fog cell (forward and control instructions module) \item manages the links status communications (management links status module ). \end{itemize}  \end{minipage}    \\ \hline    
		
		\cite{huang2018v2v}  &\begin{minipage}[t] {0.5\textwidth} \begin{itemize} \item stores contextual information of vehicles \item computes the V2V path \item controlls the switching for V2V offloading \end{itemize}  \end{minipage}\\ \hline 
		
		\cite{JLiu2017}       &\begin{minipage}[t] {0.5\textwidth} \begin{itemize} \item updates current storage requested, location and link \item acquires and predicts system \item executes algorithm and determine optimal decision state  \item sends flow control information and decision policies state to all edge nodes   \end{itemize} \end{minipage}    \\ \hline
		
	\end{tabular}
\end{table*}

\begin{table*}[!h]
	\centering
	\caption{Functionalities of NFV in VEC}
	\label{nfv-table}
	\begin{tabular}{|c|l|} \hline
		\textbf{Related Works}  &\textbf{Functions of Network Function Virtualisation}                                                                                                                                                    \\ \hline
		
		\cite{peng2019spectrum}       &\begin{minipage}[t] {0.5\textwidth} \begin{itemize} \item manage resources centrally and dynamically \item adjust transmit power of wireless router  \item execute spectrum slicing at edge nodes   \end{itemize} \end{minipage}    \\ \hline
		
		\cite{WZhang2017}  &\begin{minipage}[t] {0.5\textwidth} \begin{itemize} \item divide into several independent VMs \item control and adjust size of a task assigned to each VM and also its processing rate\end{itemize} \end{minipage}    \\ \hline
		
		\cite{peng2018sdn}  &\begin{minipage}[t] {0.5\textwidth} \begin{itemize} \item task balancing for computation and storage between MEC servers \end{itemize} \end{minipage}    \\ \hline
		
		\cite{JWang2018} &\begin{minipage}[t] {0.5\textwidth} \begin{itemize} \item provides network virtualisation mapping based on function-group of vehicles (i.e. multimedia entertainment, mobile business, and location-based services)   \end{itemize} \end{minipage}    \\ \hline
		
		\cite{DAChekired2018} &\begin{minipage}[t] {0.5\textwidth} \begin{itemize} \item  creates and configures different VMs according to quantity of data offloaded  \end{itemize} \end{minipage}    \\ \hline

	\end{tabular}
\end{table*}

\subsection{Edge nodes}
\label{sec-fognodes}
The edge nodes are known for the nodes that promptly compute, store and transmit data located at distributed edge networks. In general, the vehicles, which are over utilised or faced a slow computation, offload (i.e., upload) their data to nearby edge nodes. On one hand, there is a case of contents like movies caching and delivery from the edge nodes instead of directly fetching from the centre cloud.  Table \ref{types-fognodes} presents the types of edge nodes employed in the ComOf and CachDel for VEC. The edge nodes are commonly characterised into two types which are:

\subsubsection{Stationary Edge Nodes}
SENs are the computing nodes that co-located at a cellular BS \cite{8466353}, RSUs \cite{8535091}, wireless access router \cite{qi2018vehicular} or any other stationary infrastructure. SENs are often connected  to an MEC server \cite{8094861}, VEC server \cite{Zhang2018}, VM server \cite{XChen2017}, SDN controller \cite{8466353} for managing data computation, storage and distribution in edge networks. Despite that, a pool of edge nodes can also share the computational resources and communication resources of a single MEC/VEC server \cite{peng2019spectrum}. This sharing may reduce the cost of VEC implementation rather than the former. Nevertheless, the comparison of latency performance between the two remains a question. In supporting cooperative and interoperability between fog servers, a localised coordinator server is introduced \cite{WZhang2017}.In addition to that, the SEN with the MEC server has higher computing platform, higher power consumption and more expensive rather than the Vehicular Edge Nodes (VENs). Since the network operator provides the SEN, the cost and revenue of computational resources become great attention in proposing the optimum offloading and catching mechanisms. 

\subsubsection{Vehicular Edge Nodes}
VENs are smart vehicles equipped with communication modules and OBUs with computing and low storage capabilities as the concept of Vehicle as a Resource (VaaR) \cite{CZhu_2018}. As shown in Table \ref{types-fognodes}, the types of VENs considered in the offloading and caching works are cars, UAVs \cite{8370877} or buses \cite{Pham_2019}. The mobility of fog nodes explicitly expands new opportunities, such as on-demand computing where the moving vehicles may offer ubiquitously their available computational resources, particularly at the area without any SEN, like in rural areas.  Leveraging Parked Vehicles (PV) \cite{YZhang2019} at the parking area, for examples in the airport or shopping mall, as a primary VEN for computing and caching is also a promising solution for VEC.  Also, parked vehicles integrated with fog node controller as a data centre can expand the storage capacity for improving the performance of delivery services in vehicular networks\cite{YZhang2018, YZhang2019}.  However, the key limitation is that a cluster of parked cars cannot be fully used as supplemental resources when their power are turned off.  Although VEN has a small storage capacity compared with SEN, the vehicles are likely to offer a cheaper computational cost.  Another significant issue is that not all VENs including the PV are keen to offer their computational resources to other vehicles. To solve this, a good reputation VEN is necessary to receive a token or reward based on the acquired utility. 

\begin{table*}[h]
	\centering
	\caption{VEC edge nodes}
	\label{types-fognodes}
	\begin{tabular}{|l|l|L{8cm}|}
		\hline
		Categories                            & Types of fog nodes                 & Related Works                                                                                                                                                                                                                                                                                                                                     \\ \hline
		\multirow{4}{*}{Vehicular Fog Nodes}  & Vehicles                           & \cite{8434345, 8466353, Magaia2018, sym10110594, LiWang2018, ZWang2018, JWang2018, qi2018vehicular, CZhu_2018, CZhu_Conf2018, wang2018contract, Zhou2019, QYuan2018, XChen2017, LPu2019, FSun2018, GQiao2018, YSun2019, YSunPro2018, JKang2019, ZNing2019,zhou2018joint, YHui2017, YHui2019, CYang2019,sun2019joint, luo2018cooperative} \\ \cline{2-3} 
		& Parked Vehicles                    & \cite{8522034, 8517127, ZSu2017, YZhang2018, YZhang2019}                                                                                                                                                                                                                                                                                         \\ \cline{2-3} 
		& Buses                              & \cite{ZWangPIMRC2017,ZWang2017, Pham_2019}                                                                                                                                                                                                                                                                                                      \\ \cline{2-3} 
		& UAVs                               & \cite{8370877, LZhangZhao2018}                                                                                                                                                                                                                                                                                                                   \\ \hline
		\multirow{8}{*}{Stationary Edge Nodes} & BS                                 & \cite{CZhu_2018,8466353, qi2018vehicular, wang2018contract, Zhou2019, GQiao2018, PLiu2019}                                                                                                                                                                                                                                                      \\ \cline{2-3} 
		& RSU                                & \cite{8434345,8535091, 8466353, Dai2018, Dai22018, XChen2017, GQiao2018, ZNing2019, YHui2017, YHui2019, PLiu2019, luo2018cooperative}                                                                                                                                                                                                            \\ \cline{2-3} 
		& Wireless Access Router             &\cite{qi2018vehicular, GQiao2018, ZNing2019, CYang2019}                                                                                                                                                                                                                                                                         \\ \cline{2-3} 
		& RSU with MEC/Fog server            & \cite{8094861, Magaia2018, LTan2018, 8535091, 8539726, YSun2019, YSunPro2018, JKang2019, el2019exploiting, JDu2019, huang2018v2v, ku2018quality, WZhang2017}                                                                                                                                                                                     \\ \cline{2-3} 
		& RSU with VEC server                & \cite{Zhang2017, Zhang2018, Dai2018, Dai22018, li2019compound,8522034}                                                                                                                                                                                                                                                                           \\ \cline{2-3} 
		& RSU with VM server                 & \cite{XChen2017}                                                                                                                                                                                                                                                                                                                                 \\ \cline{2-3} 
		& BS with MEC server                 & \cite{8434345, sym10110594, midya2018multi, CZhu_2018, CZhu_Conf2018, QYuan2018, FSun2018, zhou2018joint,li2017resource}                                                                                                                                                                                                                     \\ \cline{2-3} 
		& RSU and BS connected to MEC server & \cite{peng2018sdn, peng2019spectrum, GQiao2018}                                                                                                                                                                                                                                                                                                  \\ \hline
	\end{tabular}
\end{table*}

Another important aspect, the term of cloudlet has been used to represent the coverage and connection of VENs and SENs, such as a bus-based cloudlet \cite{ZWang2017, ZWangPIMRC2017}, a vehicle-based cloudlet \cite{ZWang2018}, roadside cloudlet \cite{midya2018multi}, cloudlet layer \cite{ZNing2019}. Similar to the cloud, a cloudlet constitutes a group of servers that are nearby to the requesting user resulting in low computation power and short communication latency \cite{eltoweissy2010towards}. If edge nodes encounter insufficient computational resources, a central cloud computing platform \cite{8434345, 8370877} is the final alternative for computation and storage that causes high computational cost, high communication overhead and a long delay. Next, we discuss the communication technologies used between edge nodes and requested vehicles.

\subsection{Communication Technologies}    
Table \ref{type-communication} summarises the underlying communication technologies from vehicles to edge nodes and cloud or vice versa. The communication technologies employed in VEC are broadly divided into vehicle-to-vehicle (V2V) \cite{LiWang2018}, vehicle-to-infrastructure (V2I) \cite{LiWang2018} or infrastructure-to-vehicle (I2V) \cite{hu2017roadside} and infrastructure-to-infrastructure (I2I) \cite{DAChekired2018}. Specifically, wireless access in
vehicular environments (WAVE) standard or also known as IEEE 802.11p \cite{ieee2010802} is an enabler for vehicles to communicate with other vehicles (i.e., V2V) on dedicated short-range communications (DSRC) frequency band \cite{luo2018cooperative, CYang2019, CZhu_2018}. In the United States, out of  75 MHz of spectrum, seven 10-MHz are assigned for channels and 5-MHz is reserved as a guard band at the 5.9 GHz frequency band under DSRC applications \cite{DSRC}. Besides that, the WAVE also supports multihop communication among vehicles. Despite DSRC, a feasible fronthaul link is assumed for the vehicle to communicate to the UAV as a fog node \cite{LZhangZhao2018}. For the parked vehicles in the building, wired Ethernet connection or available wireless hotspots can be used to establish the connections \cite{ZSu2017}. Apparently, the benefits of wired setup are faster data upload and download speed, secure and reliable connectivity, but the cost of VEC implementation might increase.

The V2I or I2V is called for communication from vehicle to RSU/BS or vice versa. The enabler for the vehicle-to-RSU (V2RSU) can be based on IEEE 802.11p \cite{DAChekired2018}, IEEE 802.22 (TV whitespace) \cite{JDu2019}, wireless local area networks (WLAN) or WiFi \cite{qi2018vehicular, peng2019spectrum}. The IEEE 802.11 or often known as DSRC is exclusively defined to support emerging Intelligent Transport System (ITS) applications in Vehicular Ad Hoc Networks (VANET). Thus, adopting other wireless alternatives, such as WLAN or cellular networks, are likely unable to address some delay-sensitive and location-dependent ITS applications. The key difference between WLAN and DSRC is that the WLAN users can only communicate with the access point after the association and authentication procedures, which required several seconds. In contrast, the exchange of data in DSRC can be performed without the association and authentication procedures. In other words, the vehicles can immediately send or receive data once they switch to the communication channel without waiting for the link establishment. Meanwhile, frequent interaction between vehicles and infrastructures using cellular networks may incur high payments for VEC customers. It seems that DSRC is the most appealing solution to support the intelligent transport system. 

Apart from the RSU, the vehicle may offload to a BS under the category of infrastructure via cellular networks, WiMAX \cite{midya2018multi}, Orthogonal Frequency-Division Multiple Access (OFDMA) \cite{Pham_2019}, Long-Term Evolution (LTE) LTE/4G \cite{qi2018vehicular}, 5G \cite{8370877}. Due to the scalability and high efficiency, the author in \cite{8370877} suggested to use 5G throughout the VEC communication from smart vehicles up to data centers. 

The I2I communication is established between infrastructures, for example, RSU-BS, RSU-Cloud, BS-Cloud, can be either via cellular networks \cite{li2017resource}, 5G \cite{luo2018cooperative} or wired \cite{JKang2019}. Several works have assumed that the autonomous vehicles consist of multiple radio access networks interfaces \cite{peng2019spectrum}, full-duplex radio \cite{LTan2018} for connecting to edge nodes. Multi communication modules of vehicles may lead to robust and reliable connectivity, particularly during handover and beneficial in rural areas. Nevertheless, power consumption and the practicality for heterogeneous vehicular communication pose some challenges in VEC.

\begin{table*}[h]
	\centering
	\caption{Communication technologies for VEC}
	\label{type-communication}
	\begin{tabular}{|l|l|} \hline
		\textbf{Communication protocols}  &\textbf{Related Works} \\ \hline
		DSRC &\cite{8434345,Zhang2018,Zhang2017, li2019compound, Dai2018, Dai22018, CZhu_2018, CZhu_Conf2018, peng2019spectrum, YHui2017, YHui2019, luo2018cooperative}  \\ \hline
		
		Full duplex radio & \cite{LTan2018} \\ \hline
		IEEE 802.22/TV White space &\cite{JDu2019} \\ \hline
		5G & \cite{8370877, sym10110594, luo2018cooperative}\\ \hline
		OFDMA & \cite{Pham_2019,CYang2019}\\ \hline
		LTE/4G & \cite{qi2018vehicular, CZhu_2018, CZhu_Conf2018, WZhang2017, peng2018sdn, sun2019joint}\\ \hline
		Wireless LAN/WIFI/IEEE 802.11p & \cite{qi2018vehicular, peng2019spectrum, WZhang2017,peng2018sdn, GQiao2018, el2019exploiting}\\ \hline
		WiMAX & \cite{midya2018multi} \\ \hline
	\end{tabular}
\end{table*}

\subsection{Vehicle Applications}
Table~\ref{v-app} lists the three categories of vehicle applications, namely critical applications (CAs), high-priority applications (HPAs) and low-priority applications (LPAs) \cite{AVE2017}. In general, the vehicle manufacturer develops the CAs which include autonomous driving and safety support applications as the primary services initiated by the vehicles. Examples of autonomous driving applications are electronic stability control, automatic braking and adaptive cruise control, and for safety support applications are collision warning, signal violation warning, and emergency vehicles \cite{gharaibeh2017smart}. HPAs are typically driving-related and optional safety-enhancing applications, such as high definition (HD) map \cite{peng2019spectrum}, held-of view enhancement \cite{7981532}, location sight recognition \cite{qi2018vehicular} and road conditions \cite{HU201727}. The high definition (HD) map encompasses the three-dimensional location of a roadmap (e.g., lane markings, signboards, crosswalks, barriers) \cite{peng2019spectrum}.
On the other hand, LPAs are non-critical types of applications, such as infotainment \cite{gharaibeh2017smart}, speech recognition \cite{8370877}, and video processing \cite{7762913} to support a driver for instructing specific tasks. Despite the low bandwidth required for the CAs, low latency is mandatory to be fulfilled in mitigating road accidents and deaths. On the contrary, LPAs services may required high bandwidth and real-time, but the QoS is not as stringent as CAs. Therefore, it is vital for each service to be associated with its latency requirement, quality constraints, and workload profiles \cite{CZhu_2018} either in a SDN or a fog controller/coordinator. 

\begin{table*}[h]
	\centering
	\caption{Categories of vehicle applications}
	\label{v-app}
	\begin{tabular}{L{4cm}L{5cm}L{2cm}L{2cm}L{2cm}L{2cm}}
		\hline
		Categories of Vehicle Applications &Examples   & Traffic Priority & Bandwidth &Latency \\ \hline
		Critical Applications (CAs)         & autonomous driving, road-safety applications         & Highest   & Low    &Low          \\
		High-Priority Applications (HPAs)  & Image aided navigation, social-based application (i.e. intervehicle), parking navigation system, information services, traffic management,  optional safety applications & High    &Low to medium  & Low to medium     \\
		Low-Priority Applications (LPAs)   & Infotainment, multimedia, speech processing, passenger entertainment (interactive gaming) & Medium to Low  &Medium to High &Real-time       \\ \hline
	\end{tabular}
\end{table*}

Table ~\ref{sum-app} summarises the types of traffic model attempted in the VEC offloading and downloading works. A significant number of works consider either fixed or uniform distribution data \cite{ YHui2019}, and followed by a Poisson distribution traffic \cite{li2019compound}. Nevertheless, only several works employ video \cite{zhou2018joint, YHe2017, ku2018quality}, safety-sensitive messages \cite{peng2019spectrum,li2017resource, el2019exploiting}, and high definitions (HD) maps \cite{peng2019spectrum, peng2018sdn}, and real-time object recognition \cite{CZhu_2018} as well as face recognition \cite{LPu2019, Pham_2019} as the traffic model.
 
\begin{table*}[h]
	\centering
	\caption{Types of traffic used in VEC}
	\label{sum-app}
	\begin{tabular}{|l|l|}
		\cline{1-2}
		\textbf{Traffic model}               & \textbf{Related Works} \\ \hline
		Audio recoder &\cite{LPu2019} \\ \hline
		Coded packet                         &   \cite{LTan2018}          \\ \hline
		Delay-tolerant data                  &   \cite{8434345}           \\  \hline
		Elastic services                     &   \cite{8517127, FLin2018}           \\  \hline
		EV calendards (i.e. UDP) &\cite{DAChekired2018} \\  \hline
		Fixed data/Uniform distribution data &   \cite{8466353, Magaia2017,Dai2018, LiWang2018, ZWang2018, ZWang2017, ZSu2017, YHui2017, YHui2019}        \\  \hline
		HD Maps		&\cite{peng2019spectrum, peng2018sdn} \\  \hline
		Inelastic services &\cite{FLin2018} \\ \hline
		Location sight recognition service, parking lot detection  &\cite{qi2018vehicular}, \cite{LPu2019} \\ \hline 
		M/M/1 Queue                          &    \cite{Zhang2017}            \\ \hline
		Poisson Distribution                 &   \cite{7762913,8535091,8522034, AVE2017, ZWangPIMRC2017,JWang2018,WZhang2017, ZNing2019, wu2018efficient, zhou2018energy, li2019compound}           \\ \hline
		Real content			&\cite{QYuan2018} \\ \hline
		Real-time object recognition, Face recognition 		&\cite{CZhu_2018},\cite{LPu2019} \& \cite{Pham_2019} \\ \hline
		Safety-sensitive packets or messages		&\cite{peng2019spectrum, peng2018sdn, li2017resource, el2019exploiting} \\  \hline
		Socially-aware applications &\cite{LZhangZhao2018,zhou2018social} \\ \hline
		Video                                &   \cite{8370877, GQiao2018, zhou2018joint, YHe2017, ku2018quality}             \\ \hline
	\end{tabular}
\end{table*}

The authors in \cite{Zhang2017} characterised the vehicle application as \textit{M/M/1} which represented the task arrival in the MEC server. However, \cite{li2019compound} disagreed that the serving process at MEC servers is modeled as the \textit{M/M/c} queuing model with \textit{c} computation thread due to the Poisson distribution for the task arrival from the vehicle. The authors in \cite{li2019compound} suggested a \textit{G/M/c} model because the vehicular tasks follow a more general process rather than a Poisson distribution, which is evident in the results \cite{li2019compound}. 

The works in \cite{LTan2018, DAChekired2018, QYuan2018, zhou2018social, LZhangZhao2018} differ from others in terms of how they represent the vehicle applications. The authors in \cite{LTan2018} assumed coded packets to be cached at vehicles for the resource allocation optimisation problem. The researchers in \cite{DAChekired2018} dealt with a scheduling of electric vehicle energy demands via user datagram protocol-based calendars for both charging and discharging requests whereas \cite{QYuan2018} used time series analysis to model the content demands patterns subject to actual content request logs, which the model is called seasonal auto-regressive integrated moving average.  On the other hand, \cite{LZhangZhao2018, zhou2018social} explored the fog computing under the assumption of social-based vehicular services such as Waze, Mooveit, SocialDrive, CaravanTrack, and GeoVanet, in Social Internet of vehicles (SOIV) \cite{AMVegni2015}. 




Thus far, only little efforts \cite{peng2018sdn, peng2019spectrum, li2017resource, FLin2018,  LPu2019} investigated various traffic in the offloading and catching works. While \cite{peng2018sdn, peng2019spectrum} considered delay-sensitive (i.e., safety-related packets) and delay-tolerant traffic (i.e. downloading HD map), \cite{li2017resource} focused on HPAs (i.e.safety-related vehicular services) and LPAs. Besides that, \cite{FLin2018} examined the elastic and inelastic groups of services. The elastic group is typically tolerant of latency and bandwidth that explicitly divides into two: traditional elastic services (e.g., data transmission) and interactive elastic services (e.g., online chatting). In contrast, the inelastic group that is defined for delay-sensitive services can also be divided into two: hard real-time services (e.g., Voice-over-Internet Protocol) and soft real-time services (e.g., Video over Demand or living streaming). All these four kinds of services are considerably explored by \cite{FLin2018}. On the contrary, \cite{LPu2019} explored a hundred of vehicles concurrently run three types of applications, namely audio transcoder, face recognition and parking lot detection with the same deadline of 10-time slots, but different processing density level (i.e., 400, 2500 and 100000).

Answering what edge nodes appropriately served the vehicle applications, the vehicle must initially prioritise its applications, and itself serves the CPAs so that the latency is guaranteed unless insufficient computation arises. Meanwhile, HPAs may be offloaded to nearby VEN and SEN. The low-priority applications are likely to be computed by all edge nodes and even the core cloud platform. Note that offloading HPAs and LPAs to other edge nodes may only take place when the vehicle undergoes insufficient computational resources. Next, the mobility models considered in the VEC offloading and caching works are discussed.

\subsection{Mobility Model}
The mobility model in VEC  characterises the movements of vehicles with regards to their locations, velocity and direction over a period of time. In the mobility model also the researchers incorporated the distribution and movement of the vehicles at a specific area or region in the real world. The model is essential for demonstrating the performance of VEC nearer to the reality. In VEC, significant works used Simulation of Urban MObility (SUMO) \cite{midya2018multi, AVE2017, CZhu_2018, PLiu2019, YSun2019, DAChekired2018, el2019exploiting, boukerche2019loicen}, followed by specific vehicle speed\cite{8094861, sym10110594, Zhang2018, Dai2018, CYang2019, li2019compound}, vehicle acceleration \cite{8094861, ZWangPIMRC2017,Zhou2019, PLiu2019} and other models \cite{8434345, 8535091,LTan2018, JWang2018, 8434345}. They are detailed as follows:

\subsubsection{SUMO} 

SUMO \cite{SUMO2012} is an open source of multi-modal traffic simulation for vehicles, public transport, and pedestrians. It is a microscopic simulation where the vehicles can be modeled explicitly on the actual lanes or highways. Generally, SUMO can transform the real map often from Open Street Map (OSM) into a simplified road topology and integrated into the SUMO simulator with a specific configuration of vehicle speed or actual vehicle movements from trace file \cite{midya2018multi}. The Luxembourg SUMO Traffic scenario (LuST)\cite{LCodeca2015} is employed in \cite{AVE2017, CZhu_2018,midya2018multi} to emulate real traffic of Luxembourg divided into two regions, namely the highways and urban. Highways consist of high speed vehicles, and urban has a long inter-vehicle distance resulting in a low density of vehicles.
On the other hand, the real-world map of three different roads in China is imported to SUMO for simulations \cite{PLiu2019}. Three different velocity models, which are a constant velocity model, vehicle-following model, and traveling-time statistical model, are generated. The traffic simulator SUMO is integrated with a network simulator OMNeT++3 that enables to use the real maps from OSM for G6 Highway in Beijing \cite{YSun2019}. Other works also used SUMO to simulate a particular area in France with the vehicle speed of 30km/h and 80 km/h \cite{DAChekired2018}, Al-Ain City in the United Arab Emirates with the maximum speed of 100 km/h \cite{el2019exploiting}. Furthermore, content distribution works also use SUMO to generate the movement trajectories of vehicles in the area of Ottawa, Canada \cite{boukerche2019loicen} and San Franciso \cite{fan2019replication}, which are obtained from OSM.
In contrast, SUMO is used to simulate the mobility of the vehicles without the actual road topology, i.e., three-lanes, at the speed of 60 km/h and 80 km/h \cite{sun2019joint}.

\subsubsection{Vehicle Speed Range}
Referring to \cite{3GPPV2V, viriyasitavat2011dynamics}, the speed of the vehicles used is Gaussian distributed and is varied from 30 km/h to 60 km/h within an urban area of 400 $km^2$ with 400 distributed BS-fogs \cite{li2017resource}. The range of speed between 70 km/h and 150 km/h is used in
most of the works \cite{Zhang2018, Dai2018, CYang2019,li2019compound}, as shown in Table \ref{mobility}. In addition to that, the proposed DREAMS assumed the speed ranges 50 to 150 km/h and acceleration ranges 0.5 to 1.5 $m/s^2$ \cite{8094861}. Other speed ranges of the vehicles are also explored in VEC, such as 10m/s and 20 m/s \cite{ZWangPIMRC2017}, 2 to 20 m/s \cite{Zhou2019}, and 0 to 27.7m/s \cite{PLiu2019}. On the other hand, the data caching works investigated the average mobility of vehicle about 100 km/h \cite{wu2019low} and the low mobility between 40 km/h and 60 km/h \cite{taya2019concurrent}.

\subsubsection{Miscellaneous}

Another exciting work in \cite{sym10110594} used the trajectory data of all green taxis and limousines trip in New York City (NYC) \cite{GreenTaxi} for simulations. Similarly, vehicle trajectory prediction based on GPS and GIS big data is also examined in \cite{8535091}. The mobility of vehicles is modeled by discrete random jumps characterised by the contact time or sojourn time and the transmission frequency \cite{LTan2018}. It is assumed that the vehicle was connecting to the same vehicle edge node and RSU within the contact time. Evaluating the impact of vehicles mobility on the resources, thereby a stochastic geometry is applied to model the random vehicular networks, and the locations of nodes are generated by a Poisson point process (PPP)  \cite{JWang2018}. Another work used the Manhattan Grid as a vehicle mobility model in the proposed software-defined vehicular networks \cite{8434345}. 

\begin{table*}[!htb]
	\centering
	\caption{Mobility model or speed used in data offloading, caching and dissemination for VEC}
	\label{mobility}  
	\begin{tabular}{|L{4cm}|L{7cm}|}
		\hline
		\textbf{Mobility model/speed}  &\textbf{Related Works}            \\ \hline
		
		SUMO & Luxembourg\cite{midya2018multi, AVE2017, CZhu_2018}, China \cite{PLiu2019, YSun2019}, France \cite{DAChekired2018}, UAE \cite{el2019exploiting}, Canada \cite{boukerche2019loicen}, San Francisco \cite{boukerche2019loicen} \\ \hline
		
		Speed ranges & 	50 to 150 km/h \cite{8094861}, 50 to 120 km/h \cite{sym10110594}, 80km/h to 150km/h \cite{Zhang2018}, 80km/h to 140km/h \cite{Dai2018}, 80km/h to 120 km/h \cite{CYang2019}, 70km/h to 130km/h \cite{li2019compound},  10 m/s and 20 m/s \cite{ZWangPIMRC2017}, 2-20 m/s \cite{Zhou2019}, 0 to 27.7m/s \cite{PLiu2019}, 0.5 to 1.5 $m/s^2$\cite{8094861}       \\ \hline
		
		Manhattan Grid Model & \cite{8434345}             \\ \hline
		
		Vehicle trajectory prediction model     &\cite{8535091}          \\ \hline
		
		Discrete random jumps &\cite{LTan2018}  \\ \hline
		
		Stochastic geometry and Poisson &\cite{JWang2018} \\ \hline
		
	\end{tabular}
\end{table*}

\subsection{Computation Offloading and Content Delivery Issues}

The motivation of adopting edge computing in vehicular networks is primarily to solve the latency issue as the edge nodes are in proximity to the vehicles compared to the central cloud. The problem becomes worse when the autonomous vehicle occupies with many applications, yet it has a limited storage capacity. In VEC, a decision on what edge nodes compute which task or what vehicle is critical in meeting a low latency.  In addition to that,  a high-density vehicular network poses significant challenges on the computation and storage resources of geo-distributed edge nodes; and thus, optimum resource allocation is essential.  

\subsubsection{Data computation offloading}
In general, the MEC in non-vehicular networks, e.g., mobile cellular networks, served by MEC servers with several options to deploy, such as co-located with the BS \cite{liu2014concert}, radio network controllers \cite{neal2016mobile} and can be farther from the UE at the core network \cite{taleb2013follow}. On the other hand, considerable VEC offloading works assumed deploying the server at the RSU (see Table \ref{types-fognodes}) besides the BS. Another primary different is various kind of vehicles with specific mobility are used as edge nodes instead of mobile devices, which owing to a fast fading channel that affects the VEC performance. In terms of the communication protocol, DSRC is mostly used for the V2I communication for the case of RSU as edge nodes. Nevertheless, other cellular networks as highlighted previously are also evaluated in the VEC works. In VEC, the requesting vehicles are surrounded by several edge nodes (i.e., SEN and VEN) for the offloading decision. Identifying a reliable edge node for consistent connectivity is a big challenge because of the vehicle speed. Therefore, the edge nodes are currently characterised and selected in terms of available workload, central processing unit (CPU) processing rate, energy consumption, radio transmission rate, offloading cost, security and reputations. 

Since the requesting vehicle contains tempospatial data; unlike mobile cellular networks, another concern is how to optimally execute the offloading includes types of data whether segmented or not from the vehicle. Furthermore, optimum task allocation on the computational resources of edge nodes is essential to satisfy the stringent latency. Under the assumption of vehicles with multi-communication interfaces, a decision on what radio access networks is also an important problem, but thus far limited work highlighted this. From the perspective of VEN, the underutilised ones inevitably can serve many vehicles within their boundaries, but at certain extent a high density of vehicles, particularly at the road segments or junctions, can substantially serve as edge nodes and probably has high offloading requirement. The small segments within a region or small cloudlets may reduce the offloading complexity. However, this would lead to the issues of VEC overlapping region and interference called interoffloading and intraoffloading \cite{TWang2013}.

\subsubsection{Content Caching and Delivery}
The content caching at multiple locations near the users have been widely used in wireless networks. It is beneficial in reducing the content access delay, and at the same time increasing the content hit ratio or response time and network delivery rate. Caching popular content at small base-stations and even at the UE can be exploited to mitigate the burden of backhaul and also a high cost transmission from the macro base station. In contrast, the contents in VEC are time-varying, location-dependent and delay-constrained, such as automated driving services \cite{amadeo2016information}. In other words, the popular contents for VEC are likely short term with regards to location. Therefore, the service content must be
completely fetched within few number of VEC cloudlets or else, the quality of the service will deteriorate. It is challenging for edge nodes to optimize data transmitted through wireless networks while satisfying the content deadlines due to unbalanced traffic and different density of vehicular networks. Another issue is regarding the content placement policy to choose the optimum edge nodes for caching that leads to high cache hit ratio.
Both the dynamic topology of vehicular networks and the spatial distribution of data chunks have a great impact on content dissemination speed \cite{yuan2016space}. The underlying V2V and V2I wireless communication is important to determine an efficient content distribution.The edge nodes may use the contextual information of the vehicles within its coverage to schedule the contents to the requesting vehicles. 

\begin{figure*}[]
	\center{\includegraphics[width=0.7\textwidth]	{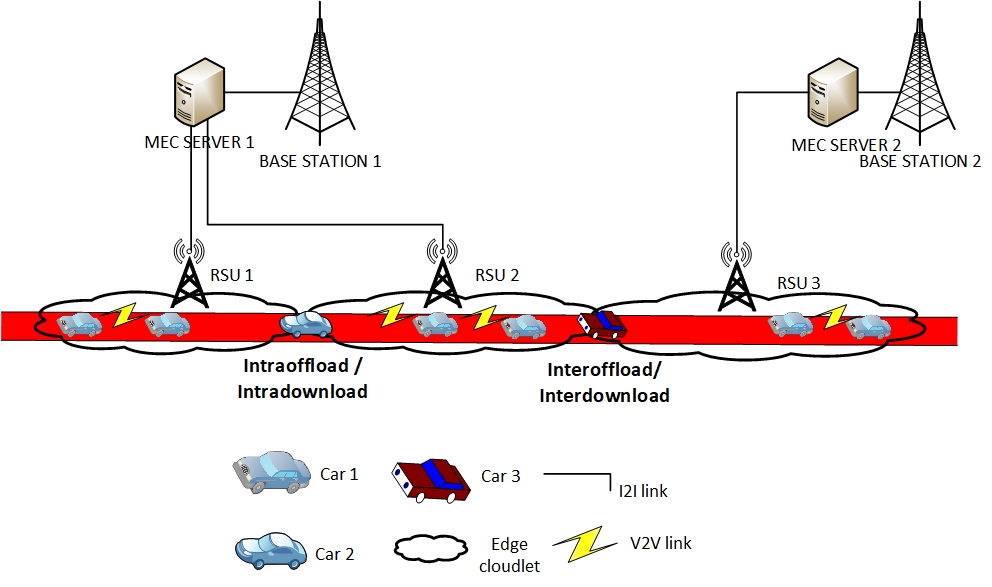}}
	\caption{Interoffloading and Intraoffloading scenario}
	\label{fig:interoffload}
\end{figure*}

\subsubsection{Service Handover}

The mobility is critical when the vehicle migrates from one fog cloudlet to another cloudlet while the data offloading or downloading is still ongoing. As depicted in Figure \ref{fig:interoffload}, this type of case is referred to intraoffloading/intradownloading in which the RSUs are connected to the same MEC. The issue becomes substantial when the vehicle handovers to an adjacent BS, as shown in the figure of the interoffloading case. The cell dwell time (i.e., cell residence time) of vehicles \cite{FSun2018} is varied and affected by several factors, such as road capacity and conditions, speed limits,  traffic lights and so forth. As highlighted previously, the unbalanced traffic and various cell dwell time lead to the resource utilisation issue of fog nodes that deserves an optimal strategy.  The moving vehicles also suffer from service interruptions due to the fast fading channel and also the availability of the fog nodes for computation in a particular area. 
%


\section{Data Computation Offloading}
\label{sec:ComOf}

\begin{figure*}[]
	\center{\includegraphics[width=0.6\textwidth]	{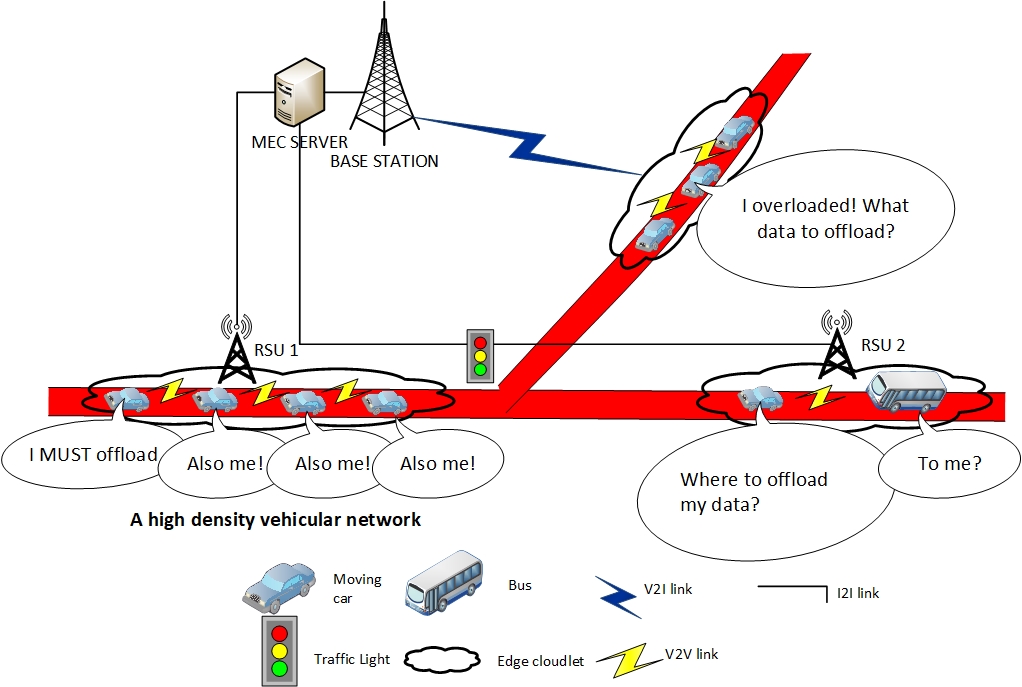}}
	\caption{Data Computation offloading scenario}
	\label{fig:offload}
\end{figure*}

The primary reason of adopting edge computing in VEC is to solve the latency issue as the edge nodes are in proximity to the requesting vehicles. Figure \ref{fig:offload} illustrates the data computation offloading scenario for non-high dense and high dense networks. Typically, the vehicles send the data to the core cloud that results in a significant delay. Considering several constraints such as the deadline, quality loss, transmission rate, energy efficiency, and fog capacity, the process of task allocation across VENs and SENs is formulated into an optimization problem, which is known as a non-deterministic polynomial-time hard (NP-hard) problem \cite{CZhu_2018} or mixed integer nonlinear programming (MINP) \cite{sym10110594, ZNing2019, Pham_2019}. The latency can be calculated based on transmission time including the round trip time between the requesting vehicle and the edge node, queuing and processing times at edge node \cite{8094861, 8535091, midya2018multi, CZhu_2018} and also handover time for the inter-offloading scenario \cite{8535091}. The existing optimisation schemes in VEC broadly emphasised several elements, such as QoS, energy, monetary, and security. Table \ref{rm-summary} summarises the characteristics of the existing offloading optimisation schemes in VEC. In this survey, the schemes are broadly divided into three groups, which are single objective optimisation, joint optimisation, and multi-objective optimisation. The three groups are discussed as follows:

\begin{table*}[]
	\centering
	\caption{Summary of resource management characteristics in Vehicular Edge Computing}
	\label{rm-summary}
	\begin{tabular}{|C{2cm}|C{1.5cm}|C{1.5cm}|C{1.5cm}|C{1cm}|C{1cm}|C{1cm}|C{1cm}|C{1cm}|C{1cm}|C{2cm}|C{1cm}|C{2cm}|}
		\hline
		\textbf{References} &\textbf{Mobility}  &\textbf{QoS-Transmission Rate}  &\textbf{QoS -Acceptance Ratio or ratio of job offloaded} & \textbf{QoS- Dateline} & \textbf{Reward System} & \textbf{Security} & \textbf{Energy Efficiency} & \textbf{Revenue} & \textbf{Cost} \\ \hline
		
		\cite{8434345}				& \checkmark		& 	    & 	    & \checkmark	     &       &  		& 		& 		& \checkmark             \\ \hline
		\cite{8094861}    			& \checkmark     &      &     & \checkmark      &       & \checkmark       &      & \checkmark     & \checkmark         \\ \hline
		\cite{8535091}    			& \checkmark     &      &      & \checkmark      &      &        & \checkmark     & \checkmark     & \checkmark          \\ \hline
		\cite{8522034}    			&      &      &      &       &       & \checkmark       &      &      & \checkmark             \\ \hline
		\cite{8517127}    			&      &      &     &       &       &        &      &      & \checkmark             \\ \hline
		\cite{8466353}    			& \checkmark     &      &      & \checkmark      &       &        &      &      &              \\ \hline
		\cite{sym10110594}			& \checkmark     &      &      & \checkmark      & \checkmark      &        &      &      & \checkmark             \\ \hline
		\cite{AVE2017}                & \checkmark     &      & \checkmark     & \checkmark      &       &        &      &      & X             \\ \hline
		\cite{Zhang2017}              & \checkmark     &      &      & \checkmark      &       &        &      &      & \checkmark             \\ \hline
		\cite{Zhang2018}              & \checkmark     &      & \checkmark     & \checkmark      &       &     & \checkmark     & \checkmark   &\checkmark     \\ \hline
		\cite{Dai2018}                & \checkmark     &      &      & \checkmark      &      &        &      &      &              \\ \hline
		\cite{Dai22018}                &      &     &      & \checkmark      &       &        &      &      &              \\ \hline
		\cite{LiWang2018}             & \checkmark     &      &      & \checkmark      & \checkmark      &        &      & \checkmark     & \checkmark     \\ \hline
		\cite{8370877}                & \checkmark     &      &      & \checkmark      &       & \checkmark       &      &      &              \\ \hline
		\cite{ZWang2018}             & \checkmark     &      & \checkmark     & \checkmark      &       &        & \checkmark     &      &              \\ \hline
		\cite{ZWang2017}              & \checkmark     &     &      & \checkmark      &       &        &      &      &              \\ \hline 
		\cite{ZWangPIMRC2017}         & \checkmark     &     & \checkmark     & \checkmark      & \checkmark      &        &      &      &        \\ \hline
		\cite{LTan2018}               & \checkmark     &      & \checkmark     &       & \checkmark      &        &      &      & \checkmark       \\ \hline
		
		\cite{JWang2018}              & \checkmark     &      & \checkmark     &      &       &        &      & \checkmark     & \checkmark       \\ \hline
		\cite{midya2018multi}         & \checkmark     &      & \checkmark     &       &       &        & \checkmark     &    &        \\ \hline
		\cite{CZhu_2018} &	&	 &\checkmark	 &	 &	 &	 &	 &	 &\\ \hline
		\cite{peng2018sdn, peng2019spectrum} &	&	 &	 &\checkmark	 &	 &	 &	 &	 &\\ \hline
		\cite{Zhou2019} &\checkmark	&	 &	 &\checkmark	 &\checkmark	 &	 &	 &	 &\\ \hline
		\cite{WZhang2017} &\checkmark	&\checkmark	 &\checkmark	 &\checkmark	 &	 &	 &\checkmark	 &	 &\\ \hline
		\cite{qi2018vehicular} &	&	 &	 &\checkmark	 & &	 &	 &	 &\\ \hline
		\cite{LZhangZhao2018} &\checkmark &\checkmark & & & & &\checkmark & & \\ \hline
		
		\cite{DAChekired2018} &\checkmark & & &\checkmark& & &\checkmark & & \\ \hline
		
		\cite{LPu2019} & & & &\checkmark &\checkmark & &\checkmark & &  \\ \hline
		\cite{FSun2018} &\checkmark & & &\checkmark & & & & & \\ \hline
		\cite{GQiao2018} &\checkmark & & &\checkmark & & & & &  \\ \hline
		\cite{YSun2019, YSunPro2018} &\checkmark & & &\checkmark & & & & &  \\ \hline
		\cite{ZNing2019} &\checkmark & & &\checkmark & & & & & \\ \hline
		\cite{zhou2018joint} & & & &\checkmark & & & & & \\ \hline
		\cite{Pham_2019} &\checkmark & & &\checkmark & & & & &\checkmark \\ \hline
		\cite{YZhang2018, YZhang2019} &\checkmark & & &\checkmark &\checkmark & &\checkmark &\checkmark &\checkmark \\ \hline
		\cite{CYang2019}  &\checkmark & & &\checkmark & & & & &\checkmark \\ \hline
		\cite{PLiu2019} &\checkmark  & & &\checkmark & & & & & \\ \hline
		\cite{SMu2018} & & & &\checkmark & & &\checkmark & & \\ \hline
		\cite{zhou2018energy} &\checkmark & & &\checkmark & & &\checkmark & & \\ \hline
		\cite{huang2018v2v} &\checkmark & & & & & & & & \\ \hline
		\cite{sun2019joint} &\checkmark & &\checkmark &\checkmark & & & & & \\ \hline
		\cite{li2017resource} &\checkmark & & &\checkmark  & & & & & \\ \hline
		\cite{el2019exploiting} &\checkmark & &\checkmark &\checkmark & & & & & \\ \hline
		\cite{JDu2019} & & & &\checkmark & & &\checkmark &\checkmark &\checkmark \\ \hline
		\cite{ku2018quality} & & & &\checkmark & & &\checkmark & & \\ \hline
		\cite{li2019compound}  &\checkmark & &\checkmark & & & & & &\checkmark \\ \hline
		
	\end{tabular}
\end{table*}

%
%
%
%
%


\subsection{Single Optimisation Approaches}

A single optimisation (SO) is defined for the computation offloading technique that maximises or minimises a single type of parameter, for instance, latency, energy or cost. Tables \ref{so} summarises the existing SO schemes in VEC and their advantages and disadvantages. The SO schemes are explained below based on their key parameters.

\subsubsection{QoS Improvement}

Significant works formulated the optimisation problem for task offloading with regards to the minimisation of latency \cite{GQiao2018, YSunPro2018, YSun2019, PLiu2019,8466353}, followed by message response time \cite{FSun2018, ZNing2019}, message overhead \cite{liu2018computation}, and the maximisation of routing and path lifetime \cite{huang2018v2v}. 

A two-stage of collaborative task offloading that comprises the vehicular function partition and task assignment stages is proposed \cite{GQiao2018}. Based on task similarity and task computing, the former classified the vehicles into two kinds of sub-cloudlets respectively, which were task offloading subcloudlet and task computing subcloudlet. Then, the later stage used the graph theory and maximum weight independent set for a two-sided matching process between the requesting vehicles and edge nodes by minimising the service latency and balancing the load of heterogeneous edge nodes. However, the results demonstrated high latency when there is a small number of resource-rich vehicles in the networks. Meanwhile, \cite{YSunPro2018} suggested an adaptive volatile upper confidence bound algorithm with load-awareness and occurrence-awareness where the utility function with regards to the classic multi-armed bandit is designed for V2V. The work is then extended by presenting an adaptive learning-based task offloading algorithm in minimising the average offloading delay \cite{YSun2019}. The algorithm improved the average delay between 10\% and 13\% compared with the former algorithm, but the scenario is only a single requesting vehicle. Another delay optimisation problem presented a pricing-based one-to-one matching and one-to-many matching algorithms for task offloading primarily at an RSU subject to minimise the total network delay \cite{PLiu2019}. The advantage is that the algorithm is validated based on three different road conditions, which are straight road, urban road with a traffic light, and winding road extracted from the realistic road topologies of Beijing and Guangdong, China. Interestingly, the lowest average delay is achieved when the offloading presents on the straight road. 

On the other hand, \cite{ZNing2019} minimised the response time for each time slot using some steps based on brand-and-bound and Edmonds-Karp algorithms. The message response time is the summation of the response time of cloudlet, the response time of parked vehicles and response time of moving vehicles as well as the delay caused by incoming messages \cite{ZNing2019}. The key advantage of the work is that the offloading is performed at both moving and parked vehicles. The findings demonstrated that the average response time achieved less than 1s with the increase of message arrival rates in Shanghai and considerably dropped when the total number of parked vehicle-based edge nodes raises. Similarly, \cite{FSun2018} devised the task scheduling optimisation problem by minimising the average response time of computation based on a modified genetic algorithm where integer coding was used. One major drawback of both works \cite{ZNing2019, FSun2018} is that the mobility analysis is not conducted. Whereas, \cite{liu2018computation} devised the computation offloading problem subject to minimise the computation overhead based on game theory, i.e., a strategic game, and was able to achieve Nash Equilibrium. The offloading problem also can be solved using total utility jobs completed at instantaneous time, and an Autonomous Vehicular Edge framework \cite{AVE2017} is proposed that comprises of four main phases, namely, job caching, discovery, ant-colony optimisation-based scheduling and data transmission. The limitation of the work is that a high computation required.
Comparing with others, the computation offloading problem can be solved also based on the longest lifetime of V2V routing path and a recovery of a broken V2V path using LifeTime-based Network State Routing and LifeTime-based Path Recovery algorithms, respectively \cite{huang2018v2v}. The results demonstrated that the average lifetime of V2V paths increased with low mobility vehicles due to stable connections. The works in \cite{liu2018computation,huang2018v2v} might be more persuasive if the authors considered the QoS and mobility.

\subsubsection{Energy Efficiency}
\cite{ZWang2018, ZWangPIMRC2017, ZWang2017, SMu2018} explored the offloading optimisation problem from the same ground to minimise the sum of energy consumption for local task execution and offloading task from the mobile device (MD). Despite that, the works are distinctive in terms of the type of edge nodes used, which are a bus-based cloudlet \cite{ZWang2017, ZWangPIMRC2017} and a vehicle-based cloudlet \cite{ZWang2018}. A vehicle-based cloudlet relaying scheme \cite{ZWang2018} allows the inter-cloudlet offloading or back to the MD if there is no cloudlet available. Due to the offloading rate performance increases with the number of buses in the cloudlet \cite{ZWangPIMRC2017, ZWang2017} as well as with the number of cloudlets \cite{ZWang2018}, the completion time is found improved for both cases. The key advantage of \cite{ZWang2018,ZWangPIMRC2017,ZWang2017} is that multi-cloudlets and high mobility are considered and yet the computation of application completion time is not properly discussed.  Another technique called Energy Consumption Minimisation (ECM) and Resource Reservation (RR) assignment for UE is proposed based on dynamic programming \cite{SMu2018}. Simulation results demonstrated that the RR assignment could achieve a near-optimal performance with proper parameter settings, and the tasks were offloaded to multiple serving vehicles via multi-radio access that beneficial in energy saving.

\subsubsection{Prices Optimisation}
The monetary involved with the price of a unit size of computing resource \cite{Zhang2018, SWang2017,JWang2018}, the price of VEC offloading time \cite{Zhang2018}, the price of energy \cite{8535091, ZWangPIMRC2017}, the price of transmission bandwidth or link \cite{JWang2018}, the incentive or reward paid to VEN \cite{LTan2018,LPu2019,Zhou2019} and the revenue generated by service provider \cite{Zhang2018, 8535091, JWang2018}. In general, monetary-based optimisation problems significantly adopted game theory, such as Stackelberg game \cite{8522034, 8535091}, a contract theoretic \cite{Zhang2018, wang2018contract, Zhou2019}, and auction \cite{YZhang2018}. Only few works presented the offloading optimisation problems based on machine learning, i.e., deep reinforcement learning \cite{qi2018vehicular, LTan2018}.

\begin{figure*}[htp]
	\center{\includegraphics[width=0.7\textwidth]	{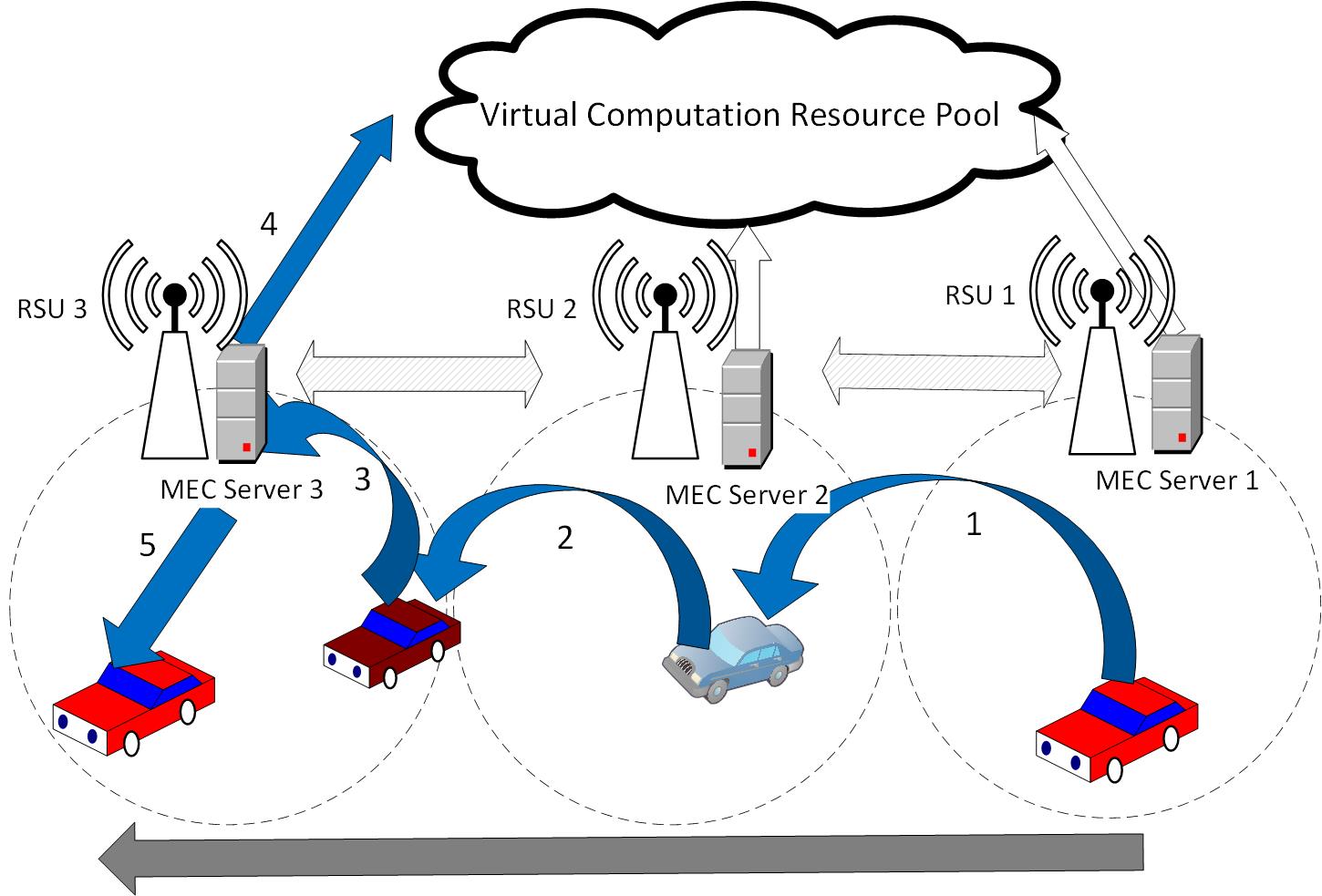}}
	\caption{A Predictive Combination-mode Offloading mechanism \cite{Zhang2017}}
	\label{fig:offloadv2}
\end{figure*}

In an optimal Predictive Combination-mode Offloading mechanism \cite{Zhang2017}, the vehicle offloaded the task file to the ahead MEC server via number of vehicle hops as illustrated in Figure \ref{fig:offloadv2}. The strength of the proposed scheme is that the transmission mode is adaptive with vehicle speeds; however, the vehicle hops may cause some delay. The results demonstrated that the mobile-edge computing servers reduced the cost of computational resources and transmission besides fast response time compared with direct vehicle-to-infrastructure (V2I), i.e., RSU with multihops backhaul RSU relay. 

Considering the minimisation of the predicted cost of the load and also the load distribution among MEC servers, a load-aware MEC offloading method called as Load-Aware Mode is proposed \cite{li2019compound}. It is seen that the proposed scheme can reduce up to 65\% of total cost and achieve approximately 100\% of the task success ratio. A mobility-aware and a location-based task offloading are introduced in \cite{CYang2019} based on Newton-Raphson methods by minimising the completed system cost includes the communication and computation costs of the required vehicle while satisfying the task latency. Numerical results showed that the proposed schemes can reduce the system costs approximately 50\% from other techniques, and the latency constraint was satisfied. However, both works \cite{li2019compound, CYang2019} overlook the types of vehicle applications that can be investigated.

Exploring a machine learning technique, specifically deep reinforcement learning \cite{qi2018vehicular} can solve the offloading problem by minimising the expected cumulative discounted reward (i.e., task execution delay) for multi-task services. The advantage of the work is that dynamic VEC environment is studied, but the results are not explicitly discussed.

\begin{table*}[!htb]
	\centering
	\caption{Single optimisation schemes in VEC}
	\label{so}
	\begin{tabular}{|C{1.5cm}|C{2cm}|L{3cm}|L{3cm}|L{3cm}|L{3cm}|}
		\hline
		{\bf Related Works}	&{\bf Optimisation Type} &{\bf Optimisation Utility}  &{\bf Optimsiation Techniques} &{\bf Advantages}  &{\bf Disadvatages}  \\ \hline

		\cite{GQiao2018} &Optimal &Minimise service latency &two-sided matching algorithm &Load balancing & High latency when small number of resource-rich vehicles\\ \hline
		
		\cite{YSun2019, YSunPro2018} &Suboptimal &Minimise average offloading delay &multi-armed bandit (MAB) & Input data size and occurrence of vehicles alert & Only a single requesting vehicle \\ \hline  	   	
		
		\cite{PLiu2019} &Optimal &Minimise total offloading delay &One-to-one and one-to-many matching & Three different mobility model and road conditions &No partial offloading at local vehicle \\ \hline
		
		\cite{liu2018computation} &Optimal & Minimise computation overhead & Game theory (i.e., strategic game)  &Computation overhead & No QoS delay and mobility \\ \hline
		
		\cite{AVE2017}    & Near-optimal     & Maximise total utility of jobs completed    & Ant Colony Optimisation (ACO) and a brute-force method   & Successful offloading  &High computation time  \\  \hline

		\cite{huang2018v2v} &Optimal &Maximum lifetime-based routing and path recovery & path routing &V2V path broken (e.g. vehicle run away) &No latency analysis and mobility \\ \hline

		\cite{FSun2018} &Optimal &Minimise average response time of computing &Modified genetic algorithm and statistical priority & Mutual dependent tasks & No mobility analysis\\ \hline
		
		\cite{ZNing2019} &Optimal &Minimise message response time  & brand-and-bound algorithm, Edmond Karps-algorithm & Offloading to both moving and parked vehicles & No mobility analysis \\ \hline  	
		
		\multirow{3}{3cm}{\cite{ZWang2018,ZWangPIMRC2017,ZWang2017}} & \multirow{3}{*}{Optimal}                   & \multirow{3}{3cm}{Minimise energy consumption for offloading }  & Sequential task graph \cite{ZWang2018}      \\ \cline{4-4}                  
		&                                            &                                                                                                                       & Exhaustive Method \cite{ZWang2017}  \\ \cline{4-4}   
		&
		&                                                                                                  & Semi-Markov Decision Process (SMDP)  and value iteration algorithm \cite{ZWangPIMRC2017}    & Multi-cloudlet and high mobility & Application completion time is uncertain \\ \hline                                           
		
		\cite{SMu2018} &Optimal &Minimise user's energy consumption for offloading &Dynamic programming & Vehicle fog node serves multiple tasks or vehicles & No specific applications and mobility  \\ \hline

		\cite{li2019compound} &optimal &Minimise the cost of offloading & Load-aware mode (LAM) &Multi vehicular MEC networks &No specific application and delay performance \\ \hline
		
		\cite{CYang2019} &Suboptimal &Minimise system costs (communication and computation) &Convex problem solved using  Newton-Raphson &Mobility and cooperation between MEC servers &No specific applications\\ \hline
		
		\cite{Zhang2017}    & Optimal  & Minimise offloading cost of both data transmission and task computation resources     & Game theory with one mixed strategies & Local and fog computing, adaptive transmission mode & Number of offloading hops may cause significant delay            \\ \hline
		
		\cite{qi2018vehicular} &Optimal & Minimise expected cumulative discounted reward & Deep reinforcement learning & Dynamic VEC environment state &No mobility\\ \hline 	
	
	\end{tabular}
\end{table*}

\subsection{Hybrid Optimisation Approaches}
Hybrid optimisation schemes are classified for the optimisation techniques proposing a combination of several mechanisms or parameters, such as joint offloading ratio and resource allocation, joint QoS and energy efficiency, and joint security with QoS. In contrast to the single optimisation, joint optimisation is often designed based on the summation of number of paramaters.

\subsubsection{Hybrid QoS}

\begin{figure*}[htp]
	\center{\includegraphics[width=0.7\textwidth]	{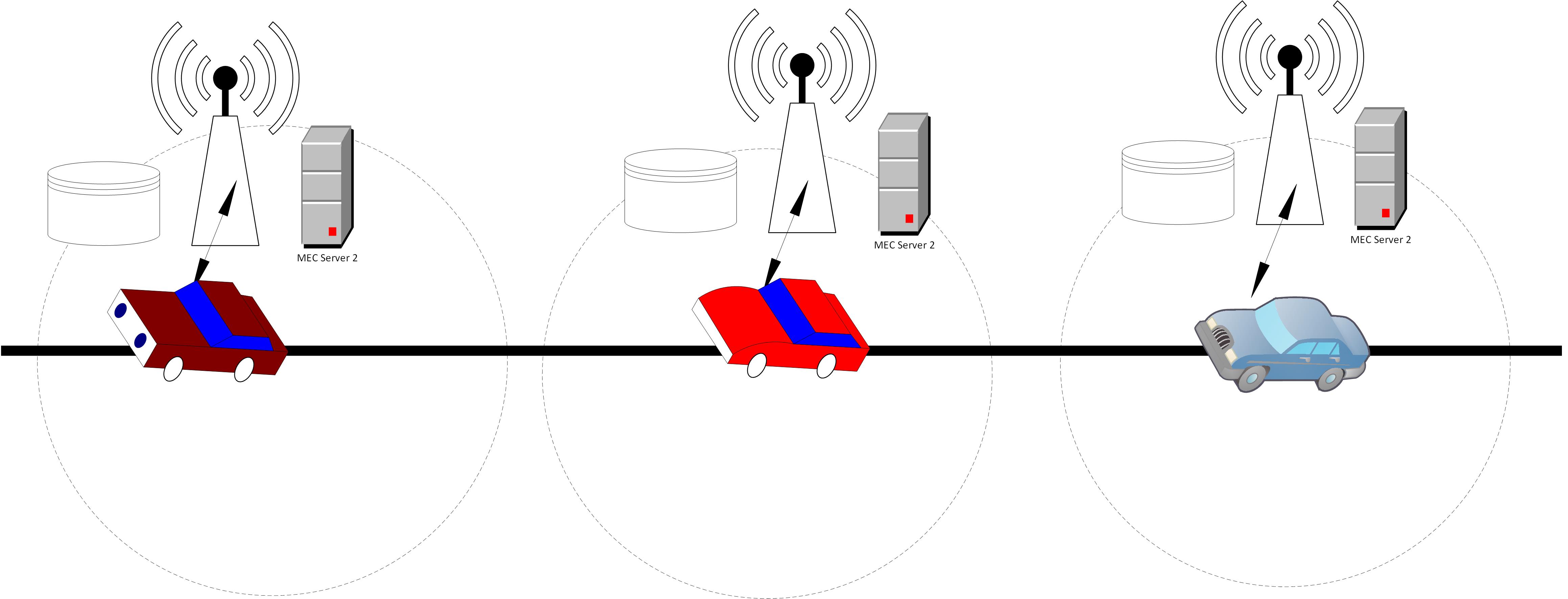}}
	\caption{System model used in Joint Optimisation for Selection, Computation and Offloading algorithm  \cite{Dai2018, Dai22018}}
	\label{fig:offloadv3}
\end{figure*}

The authors in \cite{Dai2018, Dai22018} proposed a joint optimisation for selection, computation and offloading algorithm (JSCO) to maximise QoS-based system utility. The system model consists of multiple VEC servers and multiple vehicles as illustrated in Figure \ref{fig:offloadv3}. The problem is solved using relaxation and rounding method and Lagrange multipliers. The proposed JSCO demonstrated the fair allocation of VEC servers  to the requesting vehicles besides optimum computation resource and offloading ratio. Similarly, \cite{zhou2018joint} presented joint offloading proportion and resource allocation optimisation (JOPRAO) for all vehicles to minimise the task completion time. The key difference of both works is the offloading ratio whereby \cite{Dai2018} introduced the amount of data offloads to the VEC server and computes the remaining locally,  while \cite{zhou2018joint} offloaded the sub-task between a vehicle edge node and MEC server. The offloading proportion for vehicle increases with the computation capability of the vehicle (i.e., MHz) and is enabled to achieve lower delay \cite{zhou2018joint}, but is not specifically discussed in \cite{Dai2018}. Another work \cite{peng2018sdn} formulated a network utility maximization consisted of two-level resource allocation, system bandwidth slicing, and resources partitioning for AVs.  The work is then extended to three aggregate network-wide utility maximization problems (refer to Table \ref{jo-tech}) focusing on transmit power control and solved by Linear programming relaxation and first-order Taylor series approximation and an alternate concave search algorithm \cite{peng2019spectrum}. However, both works \cite{peng2018sdn, peng2019spectrum} do not consider AVs as computing edge nodes. Unlike others, \cite{CZhu_2018} solved the offloading problem by joining two different QoS parameters, which were latency and quality task allocation using Linear Programming based Optimization and Binary Particle Swarm Optimization prior to dynamic task allocation approach \cite{CZhu_Conf2018}. The study might far more convincing if the authors had attempted the multi vehicular cloudlets.

\subsubsection{Hybrid Energy}
\cite{LZhangZhao2018} developed the total utility maximisation by jointly combined computation bits and caching bits while optimising the transmit power of each vehicle and the trajectory of UAV adaptive to the instantaneous time and environment. To solve the power optimization problem, the energy-aware dynamic power optimization problem was presented in the non-cooperation and cooperation cases under a fixed UAV trajectory. The former case was where the vehicles competed for the resources among each other, whereas the latter case was all the vehicles cooperatively share their common interest by forming a grand coalition. Simulation results demonstrated that the cooperation case with optimised trajectory achieved the best performance. However, the proposed optimisation scheme has high complexity for a real implementation. 

Meanwhile, \cite{ku2018quality} combined the energy with QoS in their proposed algorithm called QoS Loss Minimization algorithm. It comprises static RSU estimated power minimisation, Temporal Energy Balancing Algorithm and Spatial Energy Balancing Algorithm under a constraint of the delay workload. The proposed algorithm significantly reduced the QoS loss owing to power deficiency of SRSU.

The researchers in \cite{LPu2019} devised a joint control algorithm of workload assignment and VM pool resource allocation by minimising the energy consumption of vehicles for task processing. The proposed algorithm also reconciled the application latency with the incentive of vehicles in long-term. The online task scheduling algorithm, TaskSche, designed an efficient task workload assignment policy based on graph transformation and a knapsack-based VM pool resource allocation policy as the core components.

To minimise the energy consumption for the offloading, the authors in \cite{zhou2018energy} formulated a joint workload offloading and power control. The problem is solved by using an alternating direction method of multipliers (ADMM) based energy efficient resource allocation algorithm. For certain transmission power, the increase of the workload offloading portion decreased the energy consumption, and the transmission power had negative impact on the energy performance. However, the aforementioned approaches \cite{ku2018quality,LPu2019,zhou2018energy} suffer from a serious weakness in terms of vehicle mobility.
\subsubsection{Hybrid Prices}

The deep reinforcement learning approach with the multi-timescale framework is developed to solve the joint optimal resource allocation in minimising the cost of communication, caching, and computing for VEC \cite{LTan2018}. Due to the  NP-hard problem, the mobility-aware reward estimation is proposed, and yet the critical metric, such as latency is not considered for simplicity. Likewise, \cite{LiWang2018} explored a reverse auction mechanism based on Vickrey-Clarke-Groves for the cost of V2V computation offloading in which the requesting vehicles and the service provider acted as buyers and seller, respectively. Due to the fact that VCG is an optimal solution and an NP-hard problem, a suboptimal solution called the unilateral-matching-based mechanism is proposed and evaluated. The performance of sub-optimal was close to that of optimal, and the proposed sub-optimal served more vehicles with the increasing number of sellers, and the computation-intensive applications can be processed approximately by 75\% faster than local. However, the study considers the offloading to the seller first rather than the local vehicle that can lead to high payment.

Besides that, \cite{8434345} minimised the system cost, i.e., network overhead and execution time of computing task, and formulated an optimal policy called Partially Observable Markov Decision Process to select the network access, computing node and also caching node. Simulation results demonstrated that the proposed framework reduced the system cost significantly across different number of data sizes offloaded. The benefit of the proposed framework is that various edge nodes are considered from local vehicles to the cloud computing server, but the revenue to service provider is not highlighted.

Meanwhile, \cite{JDu2019} formulated two optimisation problems for Vehicle Terminals (VTs) and Mobile Radio Side Units (MRSUs) by minimising the average cost of VTs and the MRSUs in a MEC enabled IEEE 802.22-cognitive vehicular networks. The problems are solved based on Lyapunov optimisation theory with a low-complexity online algorithm called  Dual-side Dynamic Joint Task Offloading and Resource Allocation Algorithm in Vehicular Networks (DDORV), but vehicle application and mobility were not specified. The simulation results demonstrated that the DDORV achieved the highest profit for MRSU apart from the lowest cost of VTs with a trade-off the queue backlog in VTs.

Investigating a different optimisation problem, \cite{Pham_2019} minimised the total computation overhead, which was the summation of weighted-sum of task completion time and monetary cost for using cloud resources of the MEC provider. The original problem is decomposed into two-subproblem, resource allocation (RA) problem with fixed task offloading and node selection (NS) problem with the optimal-value function prior to the RA. Although the cost of both computational resources of vehicle and core cloud is considered, the bottleneck of the work is to investigate on the mobility effect.


\subsubsection{Hybrid Security} 

The reputation value of a vehicle is computed to exhibit the trustworthiness of vehicles before the offloading takes place \cite{8094861}. The computation is based on previous experience with the targeted vehicles, recommendations from the neighbouring vehicles and also from a central authority. A bargaining game is formulated for the service provider to consider the vehicle reputation value in the bargaining power and also the required task allocation, which is known as reputation-assisted optimisation. A vehicle with high reputation is given high priority to determine its required resources resulted in high user satisfaction. Despite that, the exchange of reputation messages between vehicles and VEC server might lead to a high overhead in the networks. Another similar work \cite{8370877} demonstrated data security and load balancing for the proposed 5G-enabled Intelligent Transport System (ITS) framework. It integrated the unmanned aerial vehicles (UAVs), dispatcher, edge nodes, and aggregator to maximize the processing capabilities, minimize delay, and maximize the security of the ITS. With the Bloom-filter-based security protocol used in the proposed framework, the delay can be reduced by increasing the number of edge nodes and transmitting UAV. One concern with the emergence of UAVs as edge nodes is on the battery lifetime for handling the data.

\begin{table*}[h]
	\centering
	\caption{Hybrid Optimisation in VEC}
	\label{jo-tech}
	\begin{tabular}{|C{1.5cm}|C{2cm}|L{3cm}|L{3cm}|L{3cm}|L{3cm}|}
		\hline
		{\bf Related Works}	&{\bf Optimisation} &{\bf Policy/Optimisation Problem}  &{\bf Techniques} &{\bf Advantages} &{\bf Disadvantages} \\ \hline
		
		\cite{Dai2018}, \cite{Dai22018}    & Near-optimal   & Maximise system utility of VEC server selection, offloading ratio and computation resource   & Relaxation for VEC selection, rounding method optimisation for computation resource and Lagrangian method for offloading ratio  & Offloading ratio between VEC server and local computation is considered    & High complexity algorithm is given    \\ \hline
		
		\cite{zhou2018joint} &Optimal &Minimise task completion time & Joint offloading proportion and resource allocation optimization (JOPRAO) &Offloading proportion decision and analysis &No mobility analysis \\ \hline 
		
		\cite{peng2019spectrum} &Suboptimal &Maximise network throughput with guaranteed QoS &Linear programming relaxation and first-order Taylor series approximation and an alternate concave search (ACS) algorithm &Optimal transmit power and network slicing   &No resource allocation from other vehicles \\ \hline   
		
		\cite{CZhu_2018} &Optimal and suboptimal &Minimise latency and quality loss &  Linear Programming based Optimization (LBO) and Binary Particle Swarm Optimization (BPSO) &Quality loss ratio of application   &No inter-zone or inter-cloudlets investigation \\ \hline 
		
		\cite{LZhangZhao2018} &Optimal & Maximise total utility by optimising transmit power of vehicle and UAV trajectory & dynamic programming method and search algorithm &Optimise power of vehicle   &High complexity \\ \hline
		
		\cite{ku2018quality} &Optimal &Minimise QoS loss and SRSU power consumption & QoS Loss Minimisation  &RSU energy consumption and UE association & No mobility analysis   \\ \hline
		
		\cite{zhou2018energy} &Suboptimal &Minimise energy consumption via energy-efficient workload and power control &Non-linear fractional programming  &Energy efficient offloading and power control & No mobility analysis \\ \hline
		
		\cite{LPu2019} &Optimal &Minimise energy consumption of network-wide vehicles for task processing while satisfying latency and vehicle incentive &Graph transformation and a knapsack-based VM pool resource allocation policy &Energy and incentive in proposed algorithm &No mobility analysis \\ \hline 
		
		\cite{LTan2018} & Sub-optimal      &Minimise cost of communication, storage and computation       & Deep Q-learning with multi-timescale framework (i.e. large and smale scale)  & Application-aware offloading and long term reward    &No latency analysis  \\ \hline
		
		\cite{LiWang2018}    & Optimal and suboptimal   & Maximise group efficiency, i.e., total utility offloading computation and expected cost         & Vickrey-Clarke-Groves (VCG) based reverse auction mechanism (optimal) and a unilateral-matching-based algorithm (suboptimal) & Payment procedure to seller, i.e., vehicle fog node    & Offloading to the seller first rather than local vehicle   \\ \hline

		\cite{8434345}        & Optimal     & Minimise system cost, i.e., network overhead and execution time of task computing   & Partially Observable Markov Decision Process (POMDP)  & Computing edge nodes varies from local storage,  mobile edge (i.e. MEC server)    & No revenue to service provider \\ \hline         
		
		\cite{JDu2019} &Optimal &Minimise the cost of vehicle terminals and maximise the profit of MRSU &Lyapunov optimisation &Cost and profit &No specific vehicular applications and mobility \\ \hline  
		
		\cite{Pham_2019} &Optimal & minimise the total computation overhead (i.e.  weighted-sum of task completion time and monetary cost for using cloud resources) &  Karush-Kuhn-Tucker (KKT) conditions and branch-and-bound algorithm & Cost of resources (vehicles and core cloud)  &No mobility performance \\ \hline
		
		\cite{8094861}        & Single optimal   & Reputation of vehicle and pricing policy     & Multi-weighted subjective logic and bargaining  game   & Reputation-based optimisation for vehicles & High overhead with the reputation messages \\  \hline
		
		\cite{8370877}     & Optimal     & Maximise processing capabilities (Pc), minimize delay (De) for proactive decision making, and maximize the security (Se) of the 5G enabled ITS framework & A triple Bloom filter probabilistic data structure (PDS) based scheduling technique &Three objective functions, the processing capabilities, delay and security  & Battery and storage constraints with the emergence of UAVs in handling data                                                      \\ \hline

	\end{tabular}
\end{table*}

\subsection{Multi Optimisation Approaches}

The multi-objective optimisation (MOO) is categorised for the optimisation techniques consisting of several independent utility functions for the task offloading and computing. In general, the QoS-based optimisation is separately combined with other optimisation variables, such as energy \cite{midya2018multi, WZhang2017}, and monetary \cite{JWang2018} despite other several QoS parameters \cite{sun2019joint}. The existing MOO frameworks are explained as follows.

\subsubsection{Several QoS metrics}
 The study in \cite{8466353} formulated the resource allocation with delay optimization scheme as a Markov decision process (MDP) in SDFC-VeNET, which can be solved via an equivalent Bellman equation. The solution was simplified subject to two stages, macro policy, and micro policy, where a linear approximation method and a stochastic approximation were exploited, respectively. The macro policy and micro policy handled the complete system state, i.e., traffic density state, channel state and queue state, and resource allocation, respectively. The limitation of the work is that it relied on CSI of RRH and QSI of vehicles owing to high overhead. Another work \cite{sun2019joint} designed a novel multi-objective task scheduling algorithm based on a bat algorithm that optimised two objectives functions, which were to minimise the total execution time and to maximise the total successful tasks. Extensive performance evaluation demonstrated that the proposed algorithm can reduce the task execution time and achieved a high offloading ratio as well as high number of successful tasks. However, the work does not specify the type of vehicle application used.
 
\subsubsection{QoS and Energy}
The authors in \cite{midya2018multi} developed a meta-heuristic approach called Hybrid Adaptive Particle Swamp Optimisation (PSO), which was optimised using a Genetic Algorithm (GA). In this work, every three layers computed their fitness values according to three main objectives, namely i) reduced network latency, ii) decreased energy consumption of the system, and also iii) increased availability of virtual machines. The work suffers with a serious high number of iterations for the convergence of optimal value. The authors in \cite{WZhang2017} investigated exciting work by considering intra-fog resource management in the local fog server (LFS) and inter-fog resource management in the coordinator server. In former, a convex optimization model was developed by minimising the expected total energy consumption of the fog server while satisfying the data processing rate.
On the other hand, for the inter-fog resource management, the optimal traffic was derived by minimising the maximum delay time of all fog servers and migrated massive data to nearby fog server (i.e., min-max optimisation). However, the work has dealt with fog servers and not with vehicles as edge nodes.

\subsubsection{QoS and Monetary}
 The work in \cite{8517127} used minimal processing time delay as an initial stage to solve the optimal payment of each user for CPU, RAM, and storage space of the edge device by using the Lagrangian method. Then, the maximum utility of the VSEC was set as the second stage optimisation by using the same process, which was Lagrangian for determining the optimal resource allocation scheme. Results demonstrated that the proposed optimisation approach achieved a shorter end-to-end delay and completion time compared with existing approaches due to a low number of control messages in the network and the offloading based on the available capacity at each server. The study have been more useful if the author had considered reward for the vehicles. Other researchers in \cite{JWang2018} proposed four phases in Phasing Virtual Network Embedding algorithm, which were a function-group topology using k-core decomposition, backbone part mapping, and edge part mapping phases based on Bloom filter, and the last phase was link mapping using one of the shortest path trees called the distributed Bellman-Ford, in the SINET-based vehicular networks. The mapping process for resource allocation was based on two key objectives, which were the maximum revenue ratio and maximum acceptance ratio. Although the proposed algorithm outperforms in terms of the revenue and cost performance, the QoS subject to the deadline is not yet discussed. The work in \cite{FLin2018} investigated bandwidth allocation model for TES, IES, HRTS, and SRTS and solved via a two-step approach. The first step, all the sub-optimal solutions are provided based on a Lagrangian algorithm. For the second step, the highest utility is selected as an optimal solution. The model is assumed to reduce the serving time by tuning the available bandwidth to our services without considering any constraints.

\subsubsection{Energy and Monetary Combination}

Unlike previous MOO, \cite{DAChekired2018} proposed the EV demands (i.e., charging and discharging) using a calendar policy whereby the calendars were scheduled by the appropriate fog data center based on the vehicle's location. The optimal calendar was then selected for each vehicle considering parameters, such as waiting time to plug in, the price at a specific time, distance to the EV public supply stations (EVPSS), and the demand/supply energy curve. However, the authors overlook the prices of the energy consumption in the optimisation.


\cite{8522034} proposed a Stackelberg game model between a service provider (leader) and PVs s(followers). The former was designed to minimise the overall cost of users, and the latter was to maximise the utility of PVs under the constraint of given rewards. To achieve the Stackelberg equilibrium, a subgradient-based iterative algorithm was proposed to determine the workload allocation among the PVs and concurrently minimised the overall cost of users that led to high complexity. The total amount of served vehicles reached twice compared to the existing schemes when the average workload was 2500 GHz, i.e., CPU cycles. \cite{8535091} has slightly different Stackelberg game model, whereby the game leader, i.e., RSU, issued a computing service price to vehicles within its coverage in the first stage. The utility function of the service provider is based on the total provision revenue minus the total electricity cost, which the latter is not considered in \cite{8522034}. The multi-objective optimisation is transformed into a single objective by introducing the linear weighting function. In the second stage, each vehicle as a game follower optimised its offloading proportion based on the computed service price. Then, the formulated Stackelberg game can be solved using backward induction iteratively, and the obtained strategy at each stage was shown converging to Nash equilibrium. 

Adopting a contract theoretic approach, \cite{Zhang2018} maximised the revenue of service provider by identifying the cost of computation resources and maximised the utility of vehicles based on computation resources and energy saving. Another solution is that the BS designed a contract associated with distinct performance of vehicle and rewarded the vehicles based on their vehicle types resulted in a maximum payoff \cite{wang2018contract}. However, the study seems initiated high number of contracts in the networks and the latency performance is not investigated. The work is then expanded by \cite{Zhou2019}; a contract-matching algorithm consists of two stages, which are a contract-based incentive mechanism for vehicles to promptly share their resources and a pricing-based stable matching algorithm for the assignment of UE's tasks with the preferred vehicle. The key advantage of the work is that the scenario of vehicles' private information (i.e., preference on resource sharing and the total amount of available resources) is not known at the BS, which is called as information asymmetry, is compared with that of information symmetry. On the other hand, \cite{YZhang2018} proposed a single- round multi-item parking reservation auction for two different rules, allocation rule and payment rule to determine the parking allocation and the corresponding parking payment work, respectively. The work is then extended to \cite{YZhang2019} to a multi-round multi-item parking reservation auction for the optimal offload price. The simulation revealed that with the optimal offload price, the proposed multi-round auction can improve both the profit of the fog node controller and the utility of parked vehicles. The research makes no attempt for computational resources of other edge nodes except for the parked vehicles.

\begin{table*}[!htb]
	\centering
	\caption{Multi Objective Optimisation Schemes in VEC}
	\label{mo-tech}
	\begin{tabular}{|C{1.5cm}|C{2cm}|L{3cm}|L{3cm}|L{3cm}|L{3cm}|}
		\hline
		{\bf Related Works}	&{\bf Optimisation} &{\bf Policy/Optimisation Problem}  &{\bf Techniques} &{\bf Advantages} &{\bf Disadvantages} \\ \hline		
		
		\cite{8466353}        & Optimal     & Minimise delay    & Linear approximation method and stochastic approximation method  &Macro and micro policies & Dependent on CSI of RRH and QSI of vehicles        \\ \hline

		\cite{midya2018multi} &Optimal    &Maximise fitness values at every layer: vehicular, roadside and network cloud. 	&Hybrid Particle Swamp Optimisation (HPSO)  & Offload subtasks to the VMs of the cloud that yields a maximum fitness value		& High number of iterations for the convergence of optimal value 			 \\ \hline
		
		\cite{WZhang2017} &Optimal &Minimise energy efficiency while satisfying processing data rate (i.e. intra-fog), minimise delay time while migrating massive data (inter-fog) & Convex optimisation, a min-max optimisation solved by KKT  &Resource management within a virtualised fog server and between fog servers (i.e. handover)    &No vehicular fog nodes  \\ \hline
		
		\cite{sun2019joint} &Optimal & Minimise total execution time and maximise the total weight & Bat algorithm & Number of vehicle clusters & no specific applications \\ \hline 
		
		\cite{8517127}        & Optimal   & Minimise total processing time for the optimal payment of user and maximise VSEC utility related to the optimal payment, the resource consumption, the number of devices, and the number of users & Lagrangian theory and closed form expression & Various edge nodes ranging from CPU, RAM and storage space of edge devices and the user's optimal prices for each edge resources 	&No reward for serving vehicles                   \\ \hline
		
		\cite{JWang2018} & Optimal & Maximise revenue ratio and acceptance ratio.         & Phasing Virtual Network Embedding (PVNE)  & Map or allocate virtual network onto the shared vehicular network   & Vehicle require a small database for the resource evaluation table          \\ \hline
		
		\cite{FLin2018} &Suboptimal and optimal &Maximise total utility functions of four services (TES, IES, HRTS and SRTS) &Lagrangian algorithm &Bandwidth allocation for four types of services  & No mobility analysis \\ \hline
		
		\cite{DAChekired2018} & Optimal & Minimise response time of scheduling calendars and minimise transmission delay of data &Calendar policy & Search optimum calendar for EV & no cost or pricing considered\\ \hline
		
		\cite{8535091}        & Optimal   & Maximise utility function of service provider & A two-stage of Stackelberg Game with backward induction (i.e. Nash Equilibrium) & Price of electricity for computing as the cost, and total service provision revenue     & Only RSUs as the edge node                    \\ \hline
		
		\cite{8522034}        & Optimal    & Minimises  overall cost of users and maximises utility of parked vehicles  & Stackelberg game and a subgradient-based iterative algorithm (i.e. Nash equilibrium)  & Resource allocation between parked vehicles and service provider    & High complexity                                    \\ \hline
		
		\cite{Zhang2018}        & Optimal     & Maximise the service provider (VEX provider) revenue while improving the utilities of vehicles     & A contract theoretic approach & Revenue of service provider and cost of the computation resources and energy saving   & Massive number of  contracts, no latency analysis                             \\ \hline
		
		\cite{Zhou2019} &Suboptimal & Maximise utility of base station, minimise total delay of overall networks & Contract-based incentive mechanism solved by KKT, matching based assignment & Contract design with/without information assymetry at Base Station   &High complexity   \\ \hline
		
		\cite{YZhang2018, YZhang2019} & Optimal & Maximise the payment of FNC and maximise aggregate utility of smart vehicles &Auction game , Vickrey Clarke Groves & Optimal parking allocation &Offloading to vehicles only \\ \hline
		
	\end{tabular}
\end{table*}

\section{Content Caching and Delivery}
\label{sec:CachDel}
\begin{figure*}[htp]
	\center{\includegraphics[width=0.7\textwidth]	{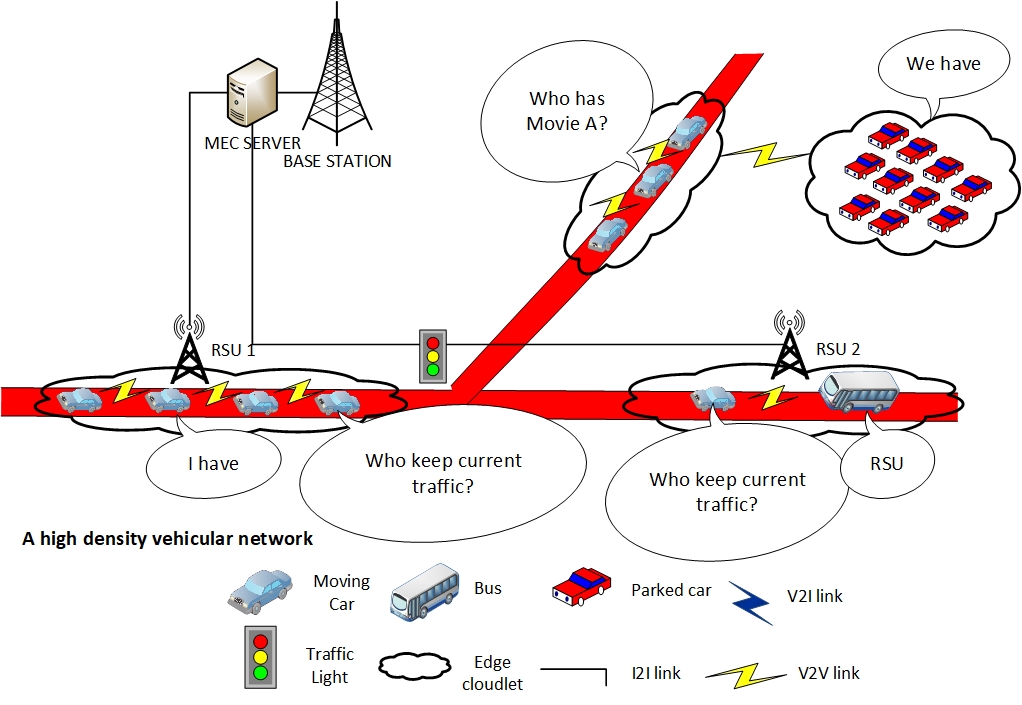}}
	\caption{Content caching and delivery}
	\label{fig:download}
\end{figure*}

Vehicular data can be characterised by three primary groups, namely location-centric, user-centric, and vehicle-centric \cite{CXu2018}. When the vehicle drives into a new city, the driver possibly acquires some spatial information on the attraction places, road conditions, live traffic, favorite restaurants or available parking spaces i.e., location-centric. Meanwhile, infotainment services like video or games may be requested by the vehicle passengers and can be analysed in terms of the users' demographics, i.e., user-centric. Information regarding the vehicles, for instance car safety, road tax, and car service or built-in sensors also can be cached at specific storage. Relaxing the burden of cloud computing, some data, such as location-centric and user-centric, can be cached locally via RSUs, vehicles or edge servers and timely shared with other vehicles, as depicted in Figure \ref{fig:download}. With the advent of IoV \cite{zhou2018social} and VSNs \cite{ZSu2016} have led to the V2V caching and communications become a reality. Caching mobile contents at the edge of networks may reduce the backhaul congestion and achieve the peak traffic apart from the lower latency \cite{fadlallah2017coding}. Nevertheless, the high variability in vehicular connectivity and rapid changes of the vehicular network topology pose some challenges on data safety and accuracy. 

In retrieving the vehicular contents, the Information-Centric Networking (ICN) \cite{khan2016saving} used the content name despite the IP address of the caching node and is prominently applied in the vehicular networks\cite{modesto2016novel,lopez2017internet}. It brings certain benefits, such as reducing the response time and the overwhelming access on the content provider. In addition to that, the key feature of the ICN is to store the most popular contents as a priority\cite{khan2016saving}. 

Table \ref{sum-datacaching} summarises the characteristics of data caching and dissemination approaches in VEC. For convenience,the data caching and dissemination approaches are classified into two, which are homogeneous cache and heterogeneous cache. Table \ref{cc-schemes} lists the content caching and dissemination schemes together with their advantages and disadvantages. The following explains the details of the data cache and dissemination mechanisms.

\begin{table*}[htp]
	\centering
	\caption{Summary of Data Caching and Dissemination Approaches in VEC}
	\label{sum-datacaching}
	\begin{tabular}{|C{2cm}|C{1.5cm}|C{1.5cm}|C{1.5cm}|C{1cm}|C{1cm}|C{1cm}|C{1cm}|C{1cm}|C{1cm}|C{2cm}|C{1cm}|C{2cm}|}
		\hline
		\textbf{References} &\textbf{Mobility}  &\textbf{QoS-Transmission Rate}  &\textbf{QoS -Acceptance Ratio/Hit Ratio} & \textbf{QoS-Delay} & \textbf{Reward System} & \textbf{Security} & \textbf{Energy Efficiency} & \textbf{Revenue} & \textbf{Cost} \\ \hline
		\cite{Magaia2017, Magaia2018} &     &      &      &       & \checkmark      & \checkmark       &     &      &            \\ \hline
		\cite{luo2018cooperative} & & & &\checkmark & & & & & \\ \hline
		\cite{QYuan2018} &\checkmark &	 &	 &\checkmark	 &	 &	 &	 &	 &\\ \hline
		\cite{YHui2017,YHui2019} &\checkmark & & &\checkmark & & & & &\checkmark \\ \hline
		\cite{JKang2019} &\checkmark & & & &\checkmark &\checkmark & & &  \\ \hline
		\cite{zhou2018social} &\checkmark &\checkmark & & & & & & & \\ \hline
		\cite{el2019exploiting} &\checkmark & &\checkmark &\checkmark & & & & & \\ \hline
		\cite{sym10110594}			& \checkmark     &      &      & \checkmark      & \checkmark      &        &      &      & \checkmark             \\ \hline
		\cite{ZSu2017}                &     & \checkmark     &      &      & \checkmark      &    &     & \checkmark     & \checkmark       \\ \hline
		\cite{hui2018content} & &\checkmark & & &\checkmark & & &\checkmark &\checkmark \\ \hline
		\cite{su2017next} &\checkmark & & &\checkmark & & & & & \\ \hline
		\cite{FSun2018} &\checkmark & & &\checkmark & & & & & \\ \hline
		\cite{qi2018vehicular} &	&	 & &\checkmark	 &	 &	 &	 & &\\ \hline
		\cite{SWang2017} & & & &\checkmark & & & & &\checkmark\\ \hline
		\cite{XChen2017} & & & &\checkmark & & & & & \\ \hline
		\cite{hu2017roadside} &\checkmark &\checkmark & & & & & & & \\ \hline
		\cite{wu2019low} &\checkmark &\checkmark & &\checkmark & & & & & \\ \hline
		\cite{taya2019concurrent} &\checkmark &\checkmark & & & && & &\\ \hline
		\cite{boukerche2019loicen} &\checkmark & &\checkmark &\checkmark & & && & \\ \hline
		\cite{fan2019replication} &\checkmark & &\checkmark &\checkmark & & & & & \\ \hline
		\cite{su2018edge} &\checkmark & &\checkmark &\checkmark & & & & & \\ \hline
		\cite{zhou2018dependable} &\checkmark &\checkmark & &\checkmark & & & & & \\ \hline
		
	\end{tabular}
\end{table*}

\subsection{Homogeneous Edge Nodes}
 Homegeneous cache is defined for a single type of edge node, e.g., vehicle and RSU, that caches and shares the contents. In this paper, we broadly categorised homogeneous caching appraoches into two groups, which are non-cooperative and cooperative homogenenous caching. 
\subsubsection{Non-cooperative caching}

In the non-cooperative caching, each edge node stored and disseminated the content without relying on other homogeneous edge nodes. The authors in \cite{hu2017roadside} explored a handoff decision between two RSUs for the caching service, as their model. A multi-object auction was presented in solving the RSU-caching problem with the objective was to maximise the total amount of downloaded data \cite{hu2017roadside}. The advantages of the work are that a caching-handoff mechanism and handoff delay are considered. The results demonstrated that the cached-enabled RSUs were fully utilised when the vehicle density was moderate. It is interesting to discover that the increase of data block size can reduce the total downloaded segmented data due to the wasted space of unallocated data block, and the reduction is critical for the unsegmented data. However, the study was conducted for a low mobility of 20 m/s.  Meanwhile, a joint peer discovery, power control, and channel selection problem was formulated in maximising the sum of weighted transmission rate subject to the physical-social score and the spectrum efficiency for matching the vehicles, i.e., V-TX and V-RX. \cite{zhou2018social}. When multiple content providers assigned with the same spectrum resource or content consumer that can be solved using a price rising strategy and yet merely for V2V edge solution. For edge server caching, two dynamic queuing theory-based scheduling schemes were proposed based on a probabilistic function of sending a job to a server by computing the ratio of mean response time at a server to the transient response time, and another scheduling considered a queue length of the server \cite{XChen2017}. A formal compositional method called Performance Evaluation Process Algebra was used to model the scheduling algorithms in a fog-based vehicular network. Because of the consistent queue length, the algorithm based on mean response time and transient response time outperformed in an unstable vehicular server system. However, the work is only investigated for edge servers and the detailed VEC architecture is not well-discussed.

\subsubsection{Cooperative caching}
 In the cooperative caching, the same kind of edge nodes cooperatively cache and disseminate the contents in any means.

A Robust and Distributed Reputation System (REPSYS) \cite{Magaia2017, Magaia2018} consists of three different modules, namely, the reputation module (reputation collection and evaluation modules), the trust module, and the decision module based on Bayesian classification. The nodes monitored and evaluated the neighbouring nodes' behaviours with regards to the reputation rating and trust rating besides the recommendation of other nodes for the cooperation \cite{Magaia2017}. The work is then extended to an incentive mechanism for vehicles caching and disseminating data in content-centric cellular-based vehicular delay-tolerant networks \cite{Magaia2018}. The system achieved a high percentage for the misbehaving nodes detection, but the bottleneck required long detection time. On the other hand, the classification of vehicles can also be in terms of vehicles types and storage capabilities \cite{fan2019replication}. Data dissemination in the dense Internet of Vehicles (IoV) can be abstracted as a complete graph using graph theory \cite{fan2019replication, zhu2017efficient}. Two replication-based algorithms; a deterministic algorithm and a distributed randomised algorithm, were designed under heterogeneous vehicular networks occupied with different types of vehicles with varying capabilities of storage. Interestingly, the proposed randomised algorithm improved the delivery ratio up to 80\% and the latency below 1.5 ms due to the various capabilities of vehicles.  

\begin{figure*}[htp]
	\center{\includegraphics[width=0.6\textwidth]	{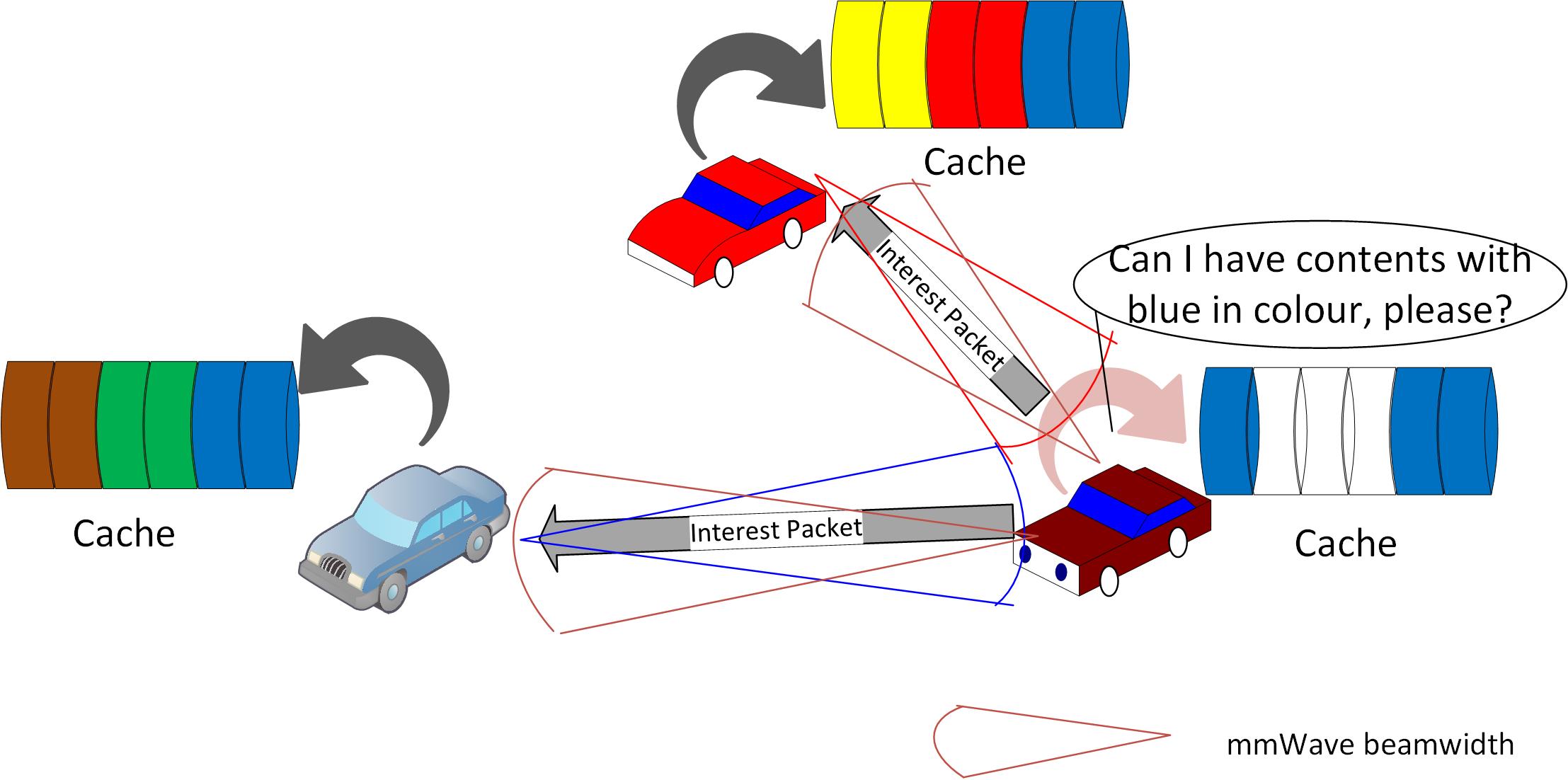}}
	\caption{ICN-based mmWave vehicular framework \cite{wu2019low}}
	\label{fig:caching}
\end{figure*}

Another cooperative caching, \cite{wu2019low} assumed that the contents are cached in vehicles' storage, and a novel ICN-based mmWave vehicular framework is proposed, which is illustrated in Figure \ref{fig:caching}.  A decentralised vehicle association algorithm called Adaptive-beamwidth Weighted Association Scheme (ABWAS) was efficiently developed to match between vehicles by maximising the content dissemination efficiency, which jointly optimised the content dissemination rate and the number of retrieval content segments. A vehicle with higher dissemination rate can support maximum transmitted content segments within the beam coherence time. Therefore, the content retrieval latency is low. The reason is that ABWAS adjusted the vehicles with their associated beamwidths in retrieving more content segments at higher transmission rate despite that high overheads.
On the other hand, the formulated joint peer discovery, spectrum allocation, and route selection optimisation problem was solved using a big data integrated coalition formation game approach via D2D-V2V multihop content distribution \cite{zhou2018dependable}. In particular, the V-RX that received the content can also cache and hence served other adjacent vehicles. The utility function proposed is to minimise the average network delay, which considered as the individual pay for each coalition member. The technique can serve approximately 90\% of vehicles on the area and achieve the average network delay below 3 ms with the increasing number of resource blocks (RBs). 

Despite that, Interest Packet transmissions were introduced \cite{boukerche2019loicen} in the proposed location-based, and information-centric (LoICen) architecture which consisted of three components, namely content request, content-location management, and content delivery. The redundant data transmission problem was solved by the vehicles' Pending Interest Table (PIT) prior to the Interest packet arrival. The data can be sent either based on location or agnostic search, i.e., link stability-based Interest forwarding (LISIC) protocol \cite{boukerche2017lisic}, to identify the vehicle that cached the required content. LoICen outperformed in terms of content delivery ratio, delay, and overhead due to the location-specific mechanism. However, the specific component that handled the LoICen is not mentioned and this is probably a BS or a coordinator server. All the previously mentioned cooperative caching methods suffer a serious limitation of only vehicles as caching nodes, i.e., V2V and not other edge nodes involved.

\subsection{Heterogenous Edge Nodes}
Heterogeneous cache involves several kinds of edge nodes that cooperatively store and disseminate the contents to the vehicles. Cooperative data caching and delivery among a variety of edge nodes are essential in the heterogeneous cache. We review the optimisation of such caching based on game theory \cite{hui2018content, YHui2017, YHui2019, ZSu2016, ZSu2017}, graph theory \cite{luo2018cooperative, QYuan2018, taya2019concurrent, qi2018vehicular}, network-based technologies and miscellaneous.

\subsubsection{Game Theory}
In general, auction game \cite{hui2018content, SWang2017}, coalition formation game \cite{YHui2017, YHui2019}, Stackelberg game \cite{ZSu2016, ZSu2017} were exploited to optimise the cost of data caching and dissemination \cite{ZSuCC2017} or jointly optimised with the transmission capabilities \cite{hui2018content, YHui2017, YHui2019}. A novel auction game model jointly considered the content dissemination requirements of an edge computing device (ECD) with regards to the transmission capability of vehicles and prices\cite{hui2018content}. A distributed two-stage relay selection algorithm was formulated for the ECD to select the optimal vehicle bids with the lowest cost in relaying the content to other vehicles using the first-price sealed-bid-auction. However, the study does not consider the content deadline. In contrast, leveraging the idle storage of parked vehicles in multiple parking areas, an iterative ascending price auction-based caching algorithm was presented \cite{SWang2017}. The resource blocks of parked vehicles were assigned to the highest bid, which was the difference between the valuation of caching and the cost paid. The process of auction might cause some delay for a high dense vehicular networks. Another work considered the cooperation among vehicles with regards to content interests (content cached in vehicles) and content requests (content needs to download) were formulated based on a coalition formation game \cite{YHui2017, YHui2019}. The selection of optimal access link was subject to the minimum cost of the content downloading time and the content price \cite{YHui2017, YHui2019}. Aforementioned game-based approaches  achieved the optimal strategy for each vehicle with a minimal cost \cite{YHui2017, YHui2019}, and yielded higher revenue to the ECDs \cite{hui2018content} than the conventional schemes. Adopting the content-centric framework \cite{ZSuCC2017}, a pricing model based on a Stackelberg game was developed for the delivery of the contents competitively from RSU or parked vehicles and cooperatively from both to the moving vehicles \cite{ZSu2017}. The cost was computed based on the computational task on a unit size of the resource (e.g., content). The proposed gradient-based iteration algorithm decreased the prices of RSU and parking area with the increasing transmission rate until the Nash Equilibrium was achieved. However, the work has a high algorithm complexity and the content deadline is not adddressed.

\subsubsection{Graph Theory}
Generally, the graph theory connects the vehicles and the edge nodes for content placement and delivery, whereby the vertices and edges can be distinguished for each proposed mechanism. The construction of the graph was assumed at the base station \cite{luo2018cooperative, QYuan2018}, RSU \cite{taya2019concurrent} and edge server \cite{qi2018vehicular}. A cooperative sharing of a large volume of vehicular data from both OBUs and RSUs was developed using an undirected neighbour graph based scheduling scheme called Balanced MWIS (BMWIS) that transformed the content distribiution problem into a MWIS problem \cite{luo2018cooperative}. The results demonstrated that the proposed BMWIS achieved the lowest average delay and a high number of served nodes for each scheduling periods. A two-level edge computing architecture was presented \cite{QYuan2018} whereby a contact graph was constructed for the content placement and solved using a tree-based heuristic method. Meanwhile, the approximation method was used to address the conflict graph for a cooperative content sharing between vehicles. Another technique based on periodic location reports from vehicles where the edge server  constructed a contact graph representing the links between the vehicles for a content placement solution \cite{qi2018vehicular}. A vehicle with a substantial gain that was proportional to the urgency of the content had a priority of broadcasting on the time slot. The implementation of graph theory-based caching might have a shortcoming of additional processing at the BS or RSU.

Despite that, mmWave data sharing algorithms for V2V communications based on graph theory scheduling is proposed in \cite{taya2019concurrent}. A vertex weighting function represented the priority of each transmission whereby a high priority transmission was assigned to the farthest vehicle from the intersection, and a low priority was given to data near the intersection, i.e., max distance scheduling. It is because the high vehicle density around the intersection is overlapped. The work is beneficial in solving the interference between beams using an approximation method called a conflict rule to improve spatial reuse, but the research might face redundant data at the intersection. With the weight parameter, the results demonstrated that the data could be shared at a large geographical area.

\subsubsection{Network-based Protocols and Technologies}
In general, data caching and dissemination for vehicular networks can also be solved using content-centric networking \cite{ZSu2016, su2017next}, blockchain \cite{JKang2019}, and network message protocols \cite{el2019exploiting}. The use of parked vehicles for delivering the contents over VSNs based on D2D communication is highlighted \cite{ZSu2016}. With CCN, the vehicles only request the name of the required content from the parked vehicles without additional overhead and the process of content interest sending, content distribution, and content replacement is detailed in \cite{ZSu2016}. The proposed technique considerably achieved a high number of successful content transmission and the shortest download delay, but required a high algorithm complexity. Another similar work employed a group of the content-centric unit (CCU) to work with vehicles and RSUs in the proposed Content-Centric Vehicular Networks framework \cite{su2017next}. The CCU can serve one or multiple RSUs, and even one RSU can be attached to several CCUs. The priority for the contents storage is subject to the request time, the arrival rate of vehicles, and the request content distribution, i.e., Zipf, which is likely can cause high overheads for the information.  However, referring to the information the lowest priority of content can be appropriately removed. A replica of required content in a selective nearby CCU was delivered to vehicle via RSU.

\begin{figure*}[htp]
	\center{\includegraphics[width=0.8\textwidth]	{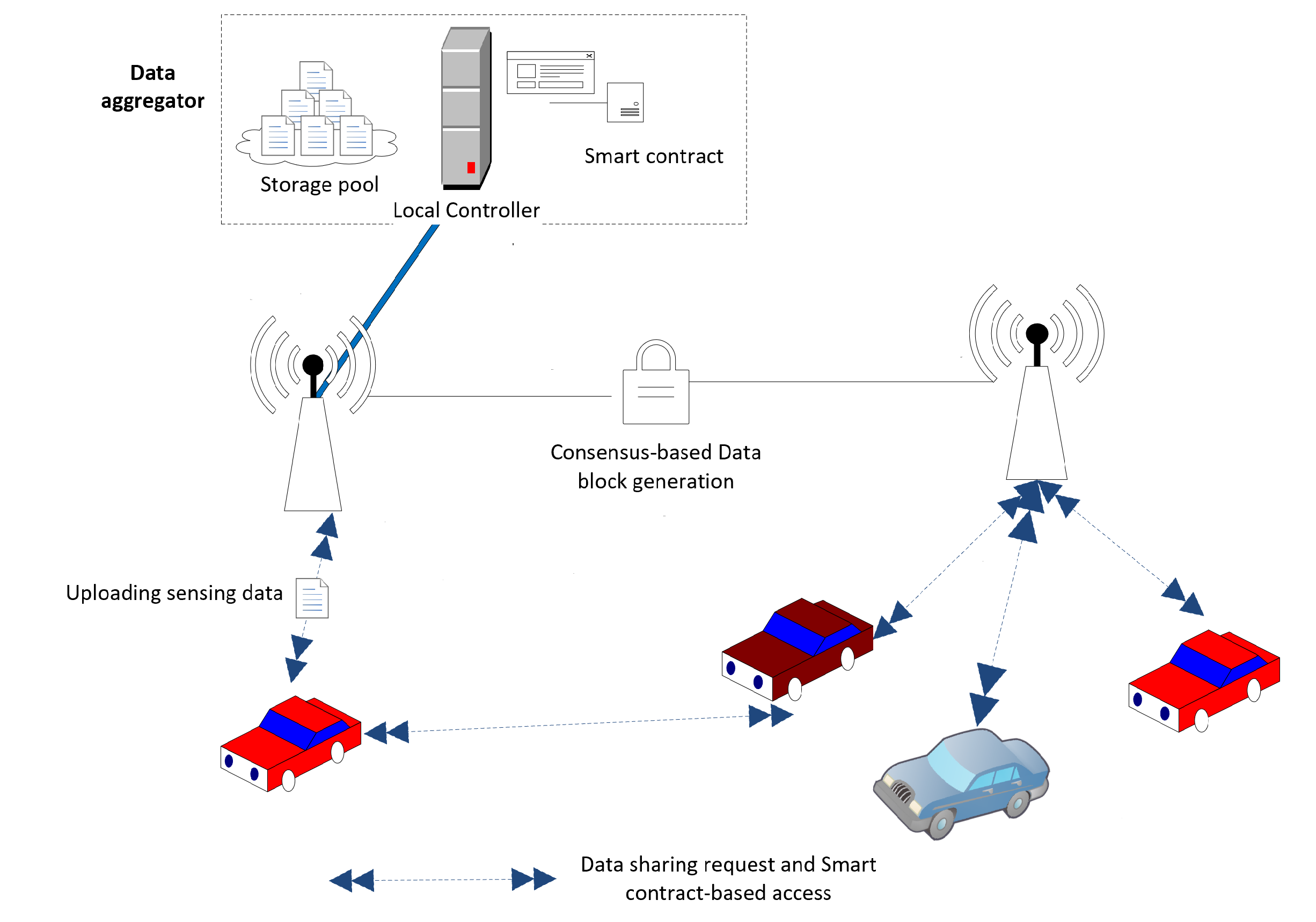}}
	\caption{Secure data storage and sharing using blockchain in VEC \cite{JKang2019}}
	\label{fig:blockchain}
\end{figure*}

Following the recent blockchain technology, Figure \ref{fig:blockchain} shows a secure data storage and sharing using blockchain in VEC \cite{JKang2019}. Smart contracts on the vehicular blockchain were proposed for a secure RSU data, and a reputation-based data sharing among vehicles called a three-weight subjective logic model for selecting the most reliable data source. The vehicle coin, which is a specific crypto-currency for vehicular edge computing and networks, is rewarded as incentives to the edge nodes in three kinds of cases: resource storage contribution, new data block update, and data providers. A local controller as seen in the figure records the total amount of contributed data storage of edgde nodes. The edge nodes (i.e., RSUs) as data aggregators periodically integrates raw data received into a data block, and requests verification from other edge nodes by broadcasting the data block.
Nevertheless, the proposed blockchain-based method will be more pervasive if QoS is taken into consideration, as in \cite{el2019exploiting}. Exploring a different mechanism, an edge-based data dissemination protocol was developed for traffic safety messages from RSUs to vehicles, and the protocol was integrated with a route request (RREQ) message and a route reply (RREP) message for delay-tolerant applications \cite{el2019exploiting}. The farthest vehicles were scheduled on the earliest timeslot for the minimal delay, similar to \cite{taya2019concurrent}. If the vehicle received a duplicate of the messages, the message broadcasting was then terminated to minimise the overheads.

\subsubsection{Miscellaneous}

Another optimisation problem was formulated to minimise the average completion time and solved using a modified genetic algorithm based joint scheduling where integer coding was used \cite{FSun2018}. The edge cloud sent a service request to the BS, which was aware of vehicular channel conditions, and hence the BS transmitted the request to the vehicle for data processing. However, a poor performance was observed for the case of high mobility cooperation.  Another caching solution introduced the utility function to maximise the hit ratio using a cross-entropy based dynamic content caching algorithm \cite{su2018edge}. The requested vehicles can fetch the content from several candidates, which are other moving vehicles, RSU, or the remote content server accordingly with regards that the average delay is well-kept. With the consideration of content popularity, content size, and also content cache capacity of the vehicle and RSU, the proposed technique achieved the highest hit ratio, the lowest relative delay, and overhead. Another work on a cost-effective resource sharing under a mixed integer nonlinear programming optimization problem was developed to minimise the cost of video service provider's in IoV \cite{sym10110594}. The target user received the video prior to the VSP optimally retrieved the video either from the moving vehicle or cloud. An incentive mechanism was introduced to encourage the vehicle sharing its contents with the BS. With a massive number of vehicles on the road, a multihop vehicle-to-vehicle connection can be performed instead of multihop backhaul relay. Apart from that, \cite{sym10110594} demonstrated that the VSP's cost was exponentially decreasing with the vehicle user's speed between 50 to 120 km/h. This is because the low speed VUs can sufficiently fetch the data from other vehicles within the delay constraint while the fast vehicle has a limited BS streaming process. However, both works in \cite{su2018edge, sym10110594} have a drawback in terms of high algorithm complexity and computation.

\begin{table*}[!htb]
	\centering
	\caption{Content Caching and Delivery Approaches in VEC}
	\label{cc-schemes}
	\begin{tabular}{|C{1cm}|C{1.5cm}|C{1.5cm}|L{3cm}|L{3cm}|L{3cm}|L{3cm}|}
		\hline
		{\bf Related Works}	&{\bf Optimisation Type} &{\bf Optimality} &{\bf Optimisation Utility}  &{\bf Optimsiation Techniques} &{\bf Advantages}  &{\bf Disadvantages}\\ \hline
		\cite{Magaia2017, Magaia2018}        &Single    &Optimal      &Maximum reputation rating   & Modified Bayesian Approach   & Vehicle's reputation and trust     & high overhead  \\ \hline		
		
		\cite{luo2018cooperative} &Single &Optimal &Maximise total number of valid content received &Graph theory &Content distribution and sharing & BS constructed the undirected neighbour graph \\ \hline
		
		\cite{QYuan2018} & Single &Suboptimal &Maximise total gain of transimission relevant to urgency of the content &Tree-based heuristic method on contact graph (content placement) and aprroximation method on conflict graph (content sharing) & Base station and autonomous vehicles for content placement   & Additional processing for content popularity \\ \hline
	
		\cite{YHui2017, YHui2019} &Single & Optimal & Minimise the cost (i.e. content downloading time and price) &Coalition formation game  & Cooperation between vehicles & High complexity  \\ \hline	
		
		\cite{JKang2019} &Single &Optimal & Maximise the final reputation of vehicle service provider & Consortium block chain and smart contract   &Data security and vehicle coin & High signalling and overheads\\ \hline
		
		\cite{zhou2018social} &Joint &Optimal &Maximize the weighted transmission rate &Pricing strategy & Power control and interference &V2V communications only \\ \hline
		
		\cite{el2019exploiting} &Single &Optimal &Minimise delay &Data dissemination protocol & Various vehicular applications &High overheads \\ \hline 
		
		\cite{sym10110594} &Single &Optimal  &Minimise VSP's cost   &Mixed nonlinear integer programming (MNIP) & Mobility analysis  & Algorithm complexity     \\ \hline
		
		\cite{ZSu2016, ZSu2017} &Multi objective &Optimal &Maximise the utility of requesting vehicles (content delivey time and conteny cost), maximise the individual utility of RSU and parked vehicles based on profits gained  &Stackelberg game &Competitive and cooperative cases between RSUs and parked vehicles & Algorithm complexity \\ \hline
		
		\cite{hui2018content} &Joint  &Optimal &Minimise bids of vehicles &Auction game &Cost and revenue & No deadline \\ \hline
		
		\cite{su2017next} &Multiple objective &Optimal & Minimise content priority and CCU distance &Information-Centric Networking (ICN) & Temporal content catched and CCU & High overheads\\ \hline
		
		\cite{FSun2018} &Single &Optimal &Minimise average response time of computing &Modified genetic algorithm and statistical priority & Cooperative task scheduling in vehicular cloud &Poor in high mobility cooperation\\ \hline
		
		\cite{qi2018vehicular} &Single &Optimal & minimise expected cumulative discounted reward (i.e. task execution delay) for all tasks & deep reinforcement & VEC environment state &  algorithm complexity  \\ \hline
		
		\cite{SWang2017} &Single &Suboptimal & Maximise the bid of content provider &Auction game & Latency, content popularity and cost & Auction process delay for resource assignment \\ \hline
		
		\cite{XChen2017} &Single &Optimal &Minimum response time &Queueing theory & Prediction of response time & Edge server only\\ \hline	
		
		\cite{hu2017roadside} &Single &Suboptimal &Maximise total downloaded data &Multi-object auction and graph theory &Segmented data and caching-specific handoff &Low mobility only 20m/s \\ \hline 
		
		\cite{wu2019low} &Joint &Optimal &Maximise content dissemination efficiency &Alpha-fair utility function \cite{perfecto2017millimeter, mo2000fair} &Vehicle associations with mmWave beam  & V2V communications only\\ \hline 
		
		\cite{taya2019concurrent} &Single &Suboptimal & Maximum weight of vertex  &Graph-based &Beam interference and priority transmission & Redundant data at intersections \\ \hline
		
		\cite{boukerche2019loicen} &Single &Optimal &Minimum defer time (latency) &LoICen &Mitigate redundant content dissemination, location-based and agnostic search & High overheads \\ \hline
		
		\cite{fan2019replication} &Single  &Optimal & Maximum gap between the vehicle values &Complete graph, i.e., graph theory &Heterogeneous vehicles &Classification of vehicle types\\ \hline
		
		\cite{su2018edge} &Single  &Optimal &Maximum hit ratio &Cross entropy based Dynamic Caching Algorithm &Content size, popularity and edge node cache size & High computation \\ \hline
		
		\cite{zhou2018dependable} &Joint &Optimal & Minimum average network delay &Coalition formation game & Channel condition and resource block & No content information\\ \hline
	\end{tabular}	 
	
\end{table*}

\section{Key Challenges, Open Issues and Future Works}
\label{sec:issues}
This section discusses the key challenges in deploying the VEC, some open issues and also potential future works.

\subsection{Key Challenges}


\subsubsection{Temporal and Spatial Vehicular Data}
Substantial vehicular application exceptional for the infotainment are time-varying spatial type of data. Such data offloading and dissemination will necessitate regular update and purge old/unpopular data in the caches. As a result, the edge nodes, particularly vehicles consume some energy and hence become drain quickly. Besides that, the high dense vehicular networks, e.g., at intersections or urban, are overlapped with number of VEC regions resulting to the flood of redundant data. Thus, the coordination for data offloading and caching must be synchronised between the overlapped areas. 

\subsubsection{Dynamic vehicular networks and unstable communication}
The mobility of vehicles leads to a dynamic vehicular network topology in which the attached RSUs, neighbouring vehicles and even the routing often varied, in particular for the case of high mobility. It is even worse when some rural areas are out of vehicular or mobile networks coverage. The vehicle edge node might equip with multiple communication modules causing to the complexity of the hardware, algorithm and definitely high battery consumption. The challenge is to sustain seamless data offloading and dissemination within the QoS requirements regardless of any circumstances and terrain. As we know, the high dense vehicular networks certainly contain with many vehicle cloudlets. The primary challenge of this is how to tackle the interference from inter-cloudlets, intra-cloudlets, and mobile networks. Another big challenge is to search for an optimum V2V communication since there is a fleet of moving vehicles around.


\subsubsection{Vehicle Cooperation}
Relying on the capability of stationary edge node, i.e., RSUs and edge servers, for computing and caching is more likely impossible since billions of connected cars are anticipated to be served. The request for content from a moving vehicle is forwarded to the RSUs when this moving vehicle enters the coverage area. When the number of requests keeps increasing, the load of an RSU to process the requests becomes heavy. Therefore, the vehicle as edge node either moving or parked could release the burden of the SEN computing. However, the challenge is that how promising the vehicle can give a full cooperation as an edge computing node. The vehicles are likely shared the computing resources to their families, relatives, friends or the person whom they know and yet not to strangers. The key challenge is to attract the vehicle owners with certain benefits for continously supporting the VEC regardless who the requestors are.

\subsection{Open Issues}

\subsubsection{Vehicular Edge Computing Architecture and Communication}

Numerous VEC architectures have been proposed with different names, edge nodes and frameworks. The underlying architecture is utmost important because the computation offloading, caching includes the resource management are solved with regards to the architectures. Thus, the VEC architecture is still an open issue that can have several options. The communication with the RSUs or edge servers using DSRC and mobile cellular technologies is significantly proposed by previous works. However, the reliable V2V communication technology is still uncertain.

\subsubsection{Data Retrieval Information and Big Data Analytics}
The ICN or VSN have been applied in the data offloading, caching and dissemination works for VEC. Adopting content-centric, location-centric or social-centric information in the edge cache are also an open issue as they have certain advantages and disadvantages. Considering the large volume, variations, temporal and spatial data features of vehicular, it is of important to classify and analyse the vehicle data effectively. The integration of big data, SDN, NFV and machine learning in the actual context of VEC remains an open issue.

\subsection{Future Works}

\subsubsection{Energy Efficiency and Various Vehicle Applications}
A number of the data offloading works addressed on the energy efficiency while only a little data caching and dissemination works highlighted this issue. Joint optimisation of energy, monetary (e.g. incentives) and QoS in providing fine-granularity of resource management in VEC particularly for both cases of computational offloading and data caching and delivery is a potential research direction. Susbtantial works explored for a single type of vehicle application and its associated QoS. A great concern must also be addressed for multiple applications and their QoS requirements.

\subsubsection{Data Security and Privacy}
The proposed framework involves a variety of data, the source of which covers from mobile devices to infrastructures. Therefore, higher data security and privacy mechanisms should be developed. The underlying security technology should be investigated to ensure a secure communication and also maintain the confidentiality of data. 
There is a serious security issue in current vehicular communication, including false information release, traffic scene forgery, and so on. Therefore, security authentication and privacy protection are the main concerns in vehicular networks. A vehicular user is necessary to recognise reliable traffic information for safe driving. Moreover, vehicular user information should be protected to prevent data breaching.


\subsubsection{Interference Coordination} 
For an effective utilisation of scarce resources, resource reuse and network density are adopted in future cellular networks. Interference would be more severe when radio behaviour occurs in densely deployed topology. Thus, the interference problem is serious in vehicular networks, especially co-channel interference. Additionally, D2D technology is applied in vehicular networks. How to deal with the interference between D2D communication and traditional communication needs to be investigated.

\subsubsection{Advancement of Vehicle Communication and Technologies Support}
In VSNs, due to similar social activities, vehicles may have the same interest for certain content and exchange between each other. However, the content is relayed by RSUs or VENs, where direct communication between vehicles cannot be efficiently provided. Investigating multiple communication protocols for V2V and V2I is significant to demonstrate the performance enhancement in particular for the cases of interoffloading and intraoffloading. Similar to mobile social users who use mobile devices to access mobile social networks, vehicles can also form a vehicular social network. Vehicles that have the same interests may exchange contents (e.g., traffic status and road conditions) with each other. A major research issue in such network is to understand the social relations between vehicles, which could be quite different from other types of social networks \cite{ZSuCC2017}. 

\subsubsection{Advancement of UAV Edge Node}
As the future research direction of vehicles, the unmanned vehicle is having a profound impact on the automotive industry and intelligent transportation systems. The maturity and development of vehicular network technology bring reliable basic support for future unmanned vehicles. However, the unmanned vehicle still faces great challenges. Some technical problems need to be resolved, such as sensitivity and accuracy of the sensors. In addition, a quantified general technical standard and an operation standard are vital for unmanned vehicles.

\section{Conclusion}
\label{sec:conclusion}
The paper has presented a comprehensive survey on computation offloading as well as data caching and delivery approaches including the optimisation techniques. We initially introduced a general architecture of VEC and then considerably discussed existing VEC architectures and frameworks, such as the VEC layers, edge nodes, communication technologies, types of vehicle applications and mobility models. The architecture is essential in designing and developing the computation offloading, data caching and dissemination techniques for VEC. The detailed reviews, findings and comparisons of existing optimisation techniques for computation offloading as well as content caching and delivery had thoroughly discussed. Finally, some key challenges, open issues and also future works were highlighted to guarantee the feasibility and reality of VEC.  

\bibliographystyle{IEEEtran}
\bibliography{referencesDatabase}{}

\end{document}